\documentclass{article}
\usepackage[final]{neurips_2021}

\usepackage{amsfonts} %
\usepackage{graphicx} %
\usepackage{natbib}  %
\usepackage{caption} %
\usepackage{algorithm}
\usepackage{listings}
\usepackage{algorithmic}
\usepackage{amsmath}
\usepackage{booktabs}
\usepackage{multirow}
\usepackage[outdir=./]{epstopdf}
\usepackage{enumitem}
\usepackage{caption}
\usepackage{subcaption}
\usepackage{subfloat}
\usepackage{newfloat}
\usepackage{graphicx}
\usepackage{svg} 
\usepackage[normalem]{ulem}
\usepackage{framed}
\usepackage{mdframed}
\usepackage{xcolor}
\usepackage{lipsum}
\usepackage{float}
\usepackage{hyperref}
\usepackage{tabularx}
\usepackage{verbatim}
\usepackage{tikz}
\usetikzlibrary{shapes, positioning}
\usetikzlibrary{intersections}

\definecolor{shadecolor}{gray}{0.9}

\usepackage{array}
\newcolumntype{L}[1]{>{\raggedright\let\newline\\\arraybackslash\hspace{0pt}}m{#1}}
\newcolumntype{C}[1]{>{\centering\let\newline  \\\arraybackslash\hspace{0pt}}m{#1}}
\newcolumntype{R}[1]{>{\raggedleft\let\newline \\\arraybackslash\hspace{0pt}}m{#1}}

\AtBeginDocument{%
  \providecommand\BibTeX{{%
    \normalfont B\kern-0.5em{\scshape i\kern-0.25em b}\kern-0.8em\TeX}}
}
\begin{document}
\title{AI Agent Behavioral Science}

\href{}{\author{%
\textbf{Lin Chen}$^{1}$ \quad 
\textbf{Yunke Zhang}$^{2}$ \quad 
\textbf{Jie Feng}$^{2}$ \quad 
\textbf{Haoye Chai}$^{2}$ \quad 
\textbf{Honglin Zhang}$^{2}$ \\ 
\textbf{Bingbing Fan}$^{2}$ \quad 
\textbf{Yibo Ma}$^{2}$ \quad 
\textbf{Shiyuan Zhang}$^{2}$ \quad 
\textbf{Nian Li}$^{2}$ \quad 
\textbf{Tianhui Liu}$^{2}$ \\
\textbf{Nicholas Sukiennik}$^{2}$ \quad 
\textbf{Keyu Zhao}$^{2}$ \quad 
\textbf{Yu Li}$^{2}$ \quad  
\textbf{Ziyi Liu}$^{2}$ \quad \\ 
\textbf{Fengli Xu}$^{2}$ \quad 
\textbf{Yong Li}$^{2}$ \\
${1}$ Department of Computer Science and Engineering, \\The Hong Kong University of Science and Technology, Hong Kong, China; \\
${2}$ Department of Electronic Engineering, BNRist,\\Tsinghua University, Beijing, China \\
\texttt{fenglixu@tsinghua.edu.cn, liyong07@tsinghua.edu.cn}}}


\maketitle

\begin{abstract}

Recent advances in large language models (LLMs) have enabled the development of AI agents that exhibit increasingly human-like behaviors, including planning, adaptation, and social dynamics across diverse, interactive, and open-ended scenarios.
These behaviors are not solely the product of the internal architectures of the underlying models, but emerge from their integration into agentic systems operating within specific contexts, where environmental factors, social cues, and interaction feedbacks shape behavior over time.
This evolution necessitates a new scientific perspective: \textbf{AI Agent Behavioral Science}. 
Rather than focusing only on internal mechanisms, this perspective emphasizes the systematic observation of behavior, design of interventions to test hypotheses, and theory-guided interpretation of how AI agents act, adapt, and interact over time.
We systematize a growing body of research across individual agent, multi-agent, and human-agent interaction settings, and further demonstrate how this perspective informs responsible AI by treating fairness, safety, interpretability, accountability, and privacy as behavioral properties. 
By unifying recent findings and laying out future directions, we position AI Agent Behavioral Science as a necessary complement to traditional model-centric approaches, providing essential tools for understanding, evaluating, and governing the real-world behavior of increasingly autonomous AI systems.


\end{abstract}

\maketitle

\section{Introduction}  







\begin{figure}[ht]
    \centering
    \includegraphics[width=\linewidth]{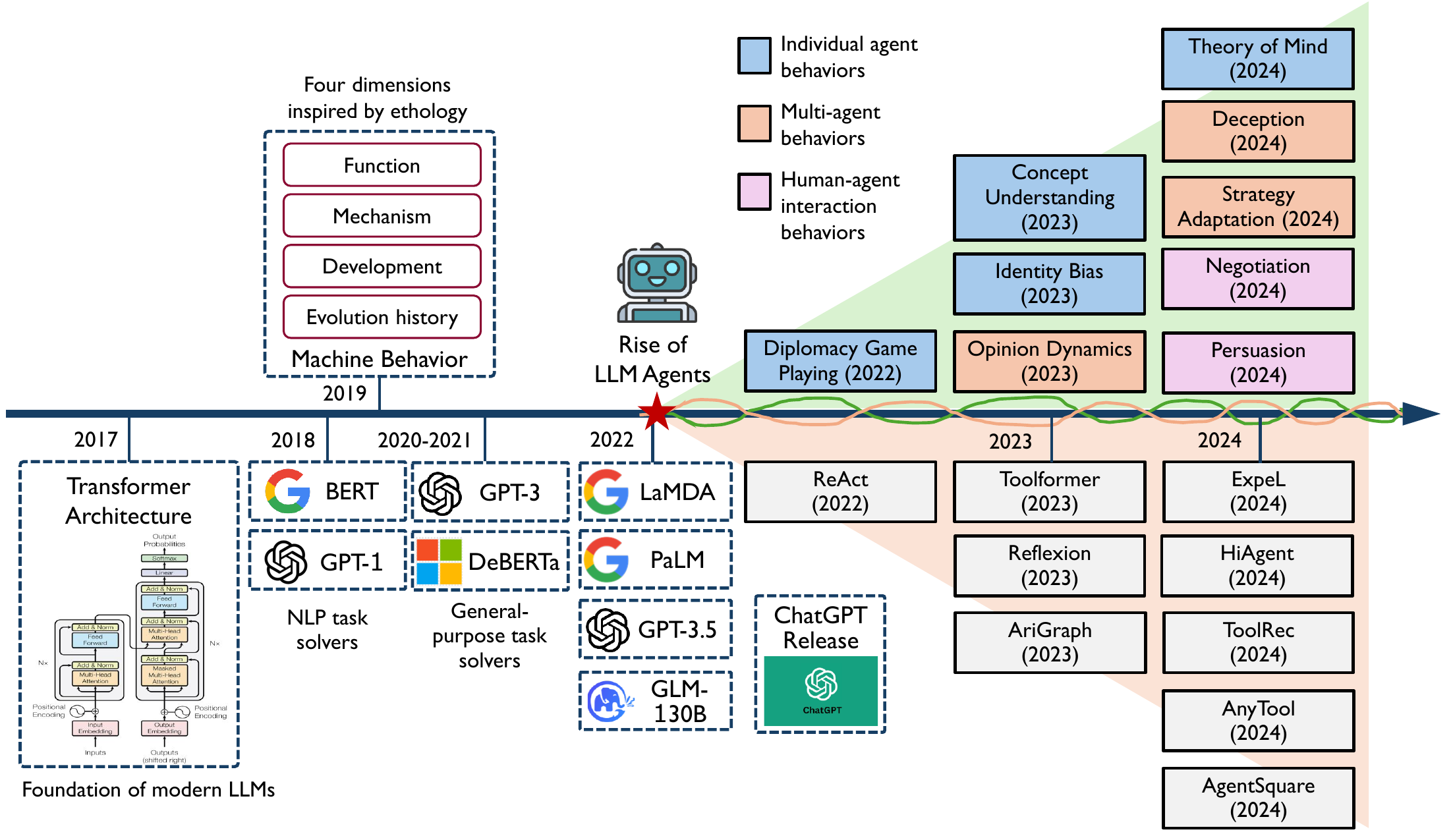}
    \caption{Development of AI technologies and understanding of AI agent behavior.}
    \label{fig:timeline}
\end{figure}

Recent advances in large language models (LLMs) have profoundly transformed how we build and interact with AI systems, particularly through the emergence of AI agents~(See Figure~\ref{fig:timeline}). 
An AI agent is an autonomous system that perceives its environment and takes actions to achieve certain goals~\cite{xi2025rise}.
For instance, when placed in a virtual village, LLM-based AI agents develop routines, hold conversations, and even organize a Valentine’s Day party~\cite{park2023generative}. 
In social deduction games like Werewolf or Avalon, they engage in deception, persuasion, and alliance formation~\cite{xu2023exploring, lan2023llm}. 
These behaviors are not pre-programmed, but emerge through situated interaction, and evolve in response to other agents, human users, and feedback from the environment. 
As such deployments proliferate, they open up a timely opportunity: to study AI systems not merely as statistical models, but as behavioral entities whose actions, adaptations, and social patterns can be empirically observed and systematically understood in context.

Traditional approaches to understanding AI have focused on internal mechanisms: architectures~\cite{nguyen2020wide}, weights~\cite{erhan2010does}, attention patterns~\cite{clark2019does}, and training objectives~\cite{radford2019language} (see Table~\ref{tab:comparison_physics_behavioral}). 
These \textit{model-centric views}, inspired by fields like physics and neuroscience, have yielded deep insights into what models encode and how they process information. 
However, they rest on the assumption that behavior can be determined and fully understood from within.
However, as AI models grow increasingly complex, pinpointing which specific components or neurons trigger particular behaviors has become challenging.
Moreover, in socially embedded and open-ended environments~\cite{krishna2022socially}, behavior is shaped not just by internal computation, but by interaction history, social context, and feedback loops. 
Model-centric tools struggle to explain the emergence of complex behaviors such as negotiation~\cite{schneider2024negotiating}, coordination~\cite{zhang2023exploring}, and deception~\cite{xu2023exploring}. 
Crucially, such behaviors rarely emerge from the AI model alone. 
Rather, they arise when AI models are embedded in agentic systems, i.e., architectures that incorporate memory, planning, tool use, and action  modules~\cite{weng2023llm}, transforming static models into dynamic, interactive entities. 
In this light, \textit{the model is to behavior what the brain is to action}: a substrate that enables but does not determine. 
Just as human behavior cannot be understood in isolation from environment and experience~\cite{clark1998being,anderson2003embodied}, AI agent behavior must be studied as a product of not only system design but also situated interaction.

We frame this emerging perspective as the \textbf{AI Agent Behavioral Science} paradigm, i.e., the study of how AI agents act, adapt, and interact in situated contexts. 
Drawing inspiration from human and animal behavioral research, this paradigm emphasizes systematic observation of behavior, hypothesis-driven intervention design, and theory-informed interpretation to uncover agent behavioral regularities and mechanisms. 
It asks not only \textit{what models can do in principle}, but \textit{what agents actually do in practice}, and more specifically, how behavioral patterns emerge, stabilize, generalize, or misalign over time given specific roles, incentives, environments, and peers. 
While much current research focuses on LLM-based agents, the core questions generalize to any AI system capable of goal-directed interaction, whether symbolic, embodied, or multimodal.
Importantly, this paradigm also unlocks new pathways for advancing responsible AI~\cite{dignum2019responsible}, reframing fairness, safety, interpretability, accountability, and privacy from static and one-shot properties of models to dynamic and context-dependent attributes.

This paradigm builds upon several foundational works.
Rahwan et al.~\cite{rahwan2019machine} call for a science of machine behavior that treats AI systems as empirical subjects of behavioral study.
Mei et al.~\cite{mei2024turing} demonstrate how behavioral science tools can be repurposed to assess LLM preferences and traits, drawing comparisons with a global dataset of human behavior.
Both scholarly comments~\cite{meng2024ai} and dedicated venues~\cite{lakkaraju2024first} have started to recognize the importance of behavioral science toward understanding and building AI agents.
While these works outline the promise of a behavioral approach, they are largely conceptual or programmatic.
By contrast, our paper takes a step further by organizing this perspective into a coherent research paradigm, systematizing emerging empirical findings, and identifying shared methods, dimensions, and open questions.
We also situate this work within broader sociotechnical conversations about AI in society.
Tsvetkova et al.~\cite{tsvetkova2024new} propose a new sociology of human-machine systems, viewing hybrid networks of people and AI agents as complex systems with emergent dynamics.
Brinkmann et al.~\cite{brinkmann2023machine} explore machine culture, emphasizing how AI systems increasingly participate in generating and transmitting cultural patterns.
These views reinforce the idea that AI systems should not only be engineered and interpreted, but also observed, evaluated, and governed as participants in social ecosystems.

In this paper, we aim to lay the groundwork for a scientific understanding of AI agent behavior. 
Our contributions are summarized below: 
\begin{itemize}
    \item We conceptualize AI Agent Behavioral Science as a coherent research paradigm—one that complements model-centric analysis by shifting the focus toward interaction, adaptation, and emergent dynamics of AI agents.
    \item We synthesize a growing body of work on AI agents to highlight how behavioral patterns can be observed, measured, theorized, and adapted across individual agent, multi-agent, and human-agent interaction scenarios. 
    \item We discuss how AI Agent Behavioral Science offers new possibilities to achieve responsible AI, for both measurement and optimization.
    \item We propose six promising research directions in this area.
\end{itemize}

For the rest of the paper, Section~\ref{sec:individual_agent}, \ref{sec:multi-agent}, and \ref{sec:human-ai} examine AI agent behavior in individual agent, multi-agent, and human-agent interaction settings, respectively.
Section~\ref{sec:adaptation} focuses on the adaptation and optimization of AI agent behavior, interpreting and understanding existing methods under the Fogg Behavior Model~\cite{fogg2009behavior}.
Section~\ref{sec:responsible_ai} applies the behavioral lens to responsible AI, highlighting how ethical principles are behaviorally measured and optimized. 
Finally, Section~\ref{sec:open_problems} outlines critical open questions and promising research directions for this emerging field.

\begin{table}[ht]
\centering
\caption{Contrasting perspectives on studying AI: the traditional model-centric view versus the emerging behavioral perspective. While the former seeks to explain models from the inside, the latter emphasizes understanding how AI agents act and adapt in context.}
\label{tab:comparison_physics_behavioral}
\resizebox{\textwidth}{!}{%
\begin{tabular}{@{}lcc@{}}
\toprule
\textbf{Dimension}     & \textbf{Model-centric Perspective}                  & \textbf{Behavioral Perspective}                           \\ \midrule
Core View of AI        & Mathematical or physical system                        & Situated behavioral agent                                 \\ \midrule
Analytical Focus       & Structure: architecture, optimization, representations & Behavior: decisions, interactions, adaptation             \\ \midrule
Methodological Tools   & Mathematics, information theory, neuroscience          & Psychology, behavioral science, sociology, economics \\ \midrule
Scientific Goal        & Explain and interpret AI model internals                  & Predict, evaluate, and shape AI behavior in context          \\ \midrule
Ontological Assumption & Models are fixed, analyzable functions                 & Agents are dynamic, contextual, and partially opaque      \\ \bottomrule
\end{tabular}%
}
\end{table}

\section{Behavioral Foundations of Individual AI Agents} \label{sec:individual_agent} 

\begin{figure}[ht]
    \centering
    \includegraphics[width=\linewidth]{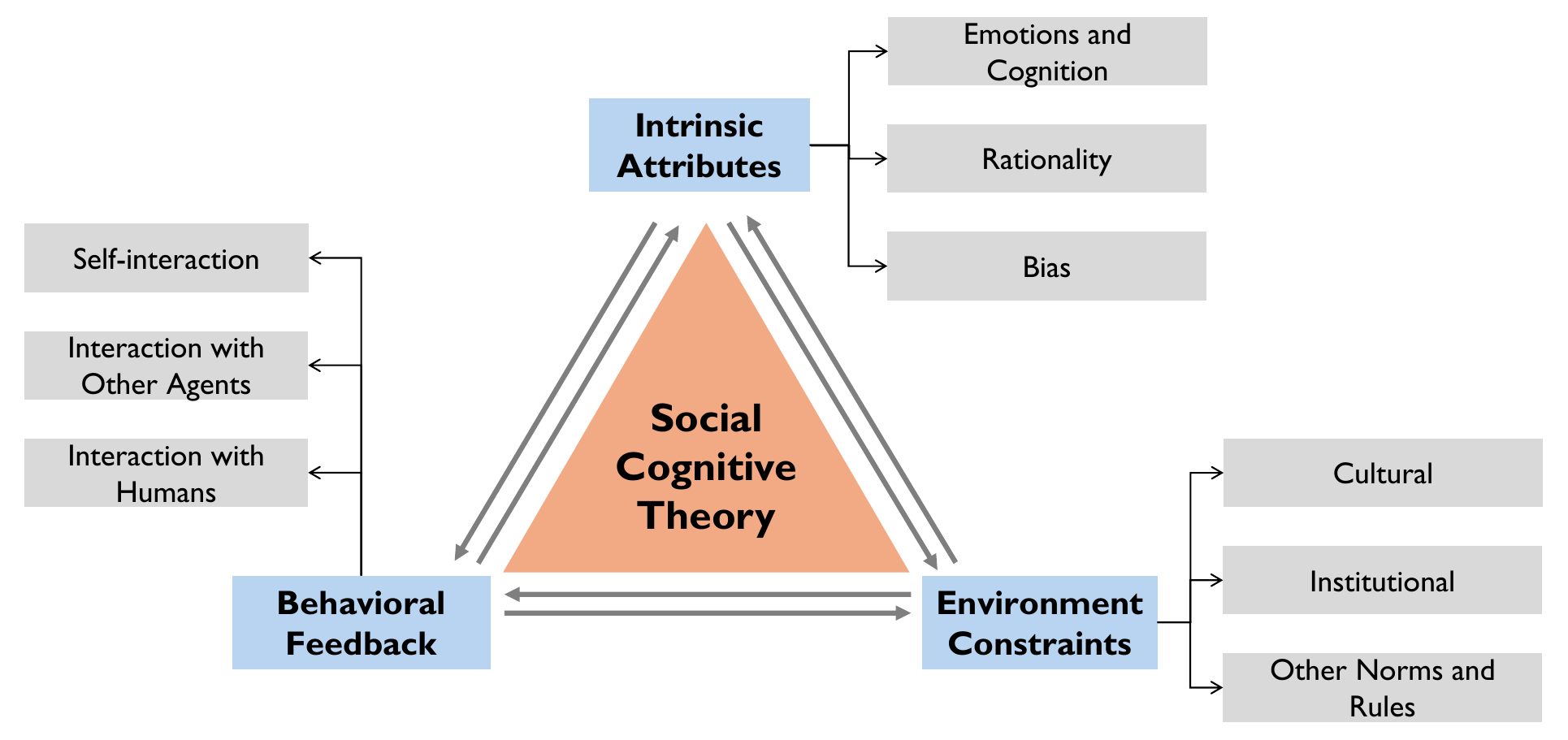}
    \caption{Determinants of individual AI agent behavior: a social cognitive perspective.}
    \label{fig:determinants_individual}
\end{figure}

Just as human individuals constitute the basic units of human societies, the behaviors of individual AI agents form the foundational layer for modeling higher-order interaction patterns and collective dynamics.
Therefore, we begin by examining the behavioral foundations at the level of individual AI agents.
Drawing inspiration from the social cognitive theory (SCT)~\cite{bandura2001social}, we organize existing research around three key dimensions that shape agent behavior over time: \textbf{intrinsic attributes}, \textbf{environmental constraints}, and \textbf{behavioral feedback} (Figure~\ref{fig:determinants_individual}). 
These dimensions offer a structured foundation for analyzing how situated decision-making and adaptation emerge in AI agents:
\begin{itemize}
    \item \textbf{Intrinsic attributes} shape agent behavior through internal traits such as emotions, cognitive patterns, value judgments, and biases, which determine how agents process information and make decisions.
    \item \textbf{Environmental constraints} influence agent behavior through external structures such as cultural norms, geographical contexts, and institutional rules, which define boundaries and social expectations.
    \item \textbf{Behavioral feedback} captures how agents continuously adapt their actions in response to social interaction, external feedback, and observed consequences.
\end{itemize}
Following this framework, we now introduce a range of emergent behaviors at the individual agent level (see Table~\ref{tab:decision-making} for a summary).

\begin{table}[ht]
\centering
\tiny
\caption{Summary of emergent individual agent behaviors.}
\label{tab:decision-making}
\resizebox{\textwidth}{!}{%
\begin{tabular}{llll} 
\toprule
\textbf{Category}         & \textbf{Topic} & \textbf{Ref.} & \textbf{Conclusion}  \\ 
\midrule
\multirow{12}{*}{\begin{tabular}[c]{@{}c@{}}Intrinsic\\Attributes\end{tabular}}    
& \multirow{5}{*}{\begin{tabular}[l]{@{}l@{}}Emotions\\and Cognition\end{tabular}}    &  \cite{le2023uncovering}   & Human-like concept understanding                     \\ 
& &  \cite{elyoseph2023chatgpt}   & Human-like emotion understanding \\
& & \cite{chen2024emotionqueen}   & Human-like emotional intelligence \\
& & \cite{strachan2024testing}   & Human-like theory of mind \\
& &  \cite{mozikov2024eai}    & Human-like decision-making influenced by emotion \\
\cmidrule(l){2-4} 
&   \multirow{4}{*}{Rationality}   &  \cite{raman2024steer}   &   Rationality emergence in large~($> 40B$) models    \\
& & \cite{xie2024different}  & Rationality varies with contexts \\
& & \cite{lore2024strategic} & Rationality varies with contexts \\
& & \cite{fan2024can} & Unsatisfactory performance in rationality \\
\cmidrule(l){2-4}
&   \multirow{3}{*}{Bias}    & \cite{gallegos2024bias}   &   Human-like bias    \\
& & \cite{acerbi2023large} & Human-like bias  \\
& & \cite{hagendorff2023human}     & ChatGPT as a turning point \\
\hline
\multirow{9}{*}{\begin{tabular}[c]{@{}c@{}}Environmental\\ Constraints\end{tabular}} 
& \multirow{4}{*}{Cultural} & \cite{myung2024blend}  & Presence of regional knowledge limitations       \\
&               & \cite{nguyen2023extracting}  & Presence of socio-cultural limitations \\
&                            & \cite{dwivedi2023eticor}  & Sensitive to regional social etiquettes \\
&                            & \cite{alkhamissi2024investigatingc}  & Cultural values alignment is achievable via prompting \\
\cmidrule(l){2-4}
& \multirow{3}{*}{Institutional} & \cite{hu2024generative}  & LLMs embody human-like social identity biases\\
&                                 & \cite{santurkar2023whose}  & LLMs hold skewed political views  \\
&                                 & \cite{yin2025safeworld}  & Achieves conformity to region bases legal norms\\
\cmidrule(l){2-4}
& \multirow{2}{*}{\begin{tabular}[l]{@{}l@{}}Other Norms\\and Rules\end{tabular}} & \cite{mozikov2024eai}  & Decision making is not affected by emotions like humans  \\
&                                       & \cite{zhao2025aligning}  & LLMs do not defend factually correct arguments when refuted \\
\hline
\multirow{8}{*}{\begin{tabular}[c]{@{}c@{}}Behavioral\\ Feedback\end{tabular}}      
&    Self-Interaction&               \cite{Silver2017MasteringTG}                  &    AI can outperform human game strategies\\
\cmidrule(l){2-4}
&                \multirow{4}{*}{\begin{tabular}[l]{@{}l@{}}Interaction with\\Other Agents\end{tabular}} &               \cite{Wu2024ShallWT}                  &                      Competing LLM agents spontaneously develop cooperative behavior\\
 & & \cite{10.5555/3295222.3295385} &AI can cooperate and deceive\\
 & & \cite{dai2024artificial} &LLM agents form cooperative societies through interaction\\
 & & \cite{Baker2019EmergentTU}&AI spontaneously learns to use tools\\
 \cmidrule(l){2-4}
&                \multirow{3}{*}{\begin{tabular}[l]{@{}l@{}}Interaction with\\Humans\end{tabular}} & \cite{mei2024turing}        &  AI adjusts behavior to framing and context\\
 & & \cite{Koster2022HumancenteredMD} &AI aligns rewards with relative contributions\\
 & & \cite{Bakhtin2022HumanlevelPI} &AI adjusts decisions by inferring players' intentions\\
\bottomrule
\end{tabular}
}
\end{table}

\subsection{Intrinsic Attributes: Intrinsic Traits and Decision Mechanisms} 


The research on the intrinsic attributes of AI agents can be divided into three main areas: (1) emotions and cognition: exploring how LLMs simulate emotional responses and cognitive processes, which are essential for enhancing their human-like interactions. (2) economic rationality: investigating how LLMs make decisions that mimic rational behavior in decision theory and game theory contexts. (3) bias: examining how LLMs may inadvertently reflect societal biases and the potential consequences of such biases on fairness and decision-making in AI applications.

\paragraph{Emotions and cognition.} 
Overall, the capabilities of GPT-4 series LLMs in this area are comparable to those of humans, at least according to the results of some standard benchmarks. 
Specifically, GPT-4's judgment of conceptual typicality is highly consistent with human judgment and far more accurate than traditional machine learning methods~\cite{le2023uncovering}. 
This task involves assessing how typical a description of something is for a given concept. 
For example, how typical is `Harry Potter' as a description of a mystery novel? CogBench~\cite{coda2024cogbench} is a more comprehensive benchmark for evaluating the psychological and cognitive abilities of LLMs, encompassing 7 psychological experiments and 10 cognitive metrics, which has been used to test 35 different LLMs. 
The results indicate that model parameter size and reinforcement fine-tuning have a significant impact on improving the cognitive abilities of LLMs and aligning their performance with that of humans. LLMs also demonstrate a relatively accurate understanding of both explicit~\cite{chen2024emotionqueen} and implicit emotions~\cite{elyoseph2023chatgpt} in human language. 
Explicit emotional recognition involves identifying emotionally charged words as labels for interpreting a text's narrative, while implicit emotional recognition involves detecting emotions hidden within events. 
For example, a story about seeing a newly opened fast-food restaurant on the way to the hospital may implicitly express sympathy for being sick. 
Based on the accurate emotional recognition by LLMs, some research has aimed at developing emotional assistants for human users to help alleviate negative emotions~\cite{qian2023harnessing}. 
For instance, at 8 p.m., a user says, "If I had a Maybach, she wouldn’t have left," and the emotional assistant might cleverly respond, "If she only rode in a Maybach, letting go wouldn’t be such a regret." 
Furthermore, many studies show that GPT-4 possesses a Theory of Mind~(ToM), meaning it has the ability to infer the mental states of others, similar to humans~\cite{strachan2024testing}. 
Furthermore, Mozikov et al.~\cite{mozikov2024eai} also suggest that emotions can influence the strategic decision-making of LLMs in a manner similar to how they affect humans, particularly in scenarios involving game playing and ethical dilemmas.

\paragraph{Rationality.} 
Raman et al.~\cite{raman2024steer} extensively test multiple LLMs on multidimensional economic rationality. 
The findings indicate that (1) LLMs with fewer than 40 billion parameters typically make random guesses for test questions; (2) GPT-4 performs most rationally; (3) self-explanation and few-shot prompting are particularly useful in enhancing LLMs' rationality. 
Nevertheless, in typical scenarios for testing economic rationality, such as game theory, the performance of the state-of-the-art model GPT-4 remains unsatisfactory~\cite{fan2024can}. 
For instance, GPT-4 sometimes fails to correctly update its beliefs based on simple factual patterns, leading to entirely unreasonable decisions. 
At the same time, research~\cite{lore2024strategic,xie2024different} also indicates that the strategic decision-making of different LLMs is affected by context to varying degrees, highlighting the issue of LLM sensitivity to prompts.

\paragraph{Bias.} 
This primarily refers to LLMs' unjust perspectives toward certain social groups~\cite{gallegos2024bias}. 
For instance, the word "whore" is disrespectful to women; the phrase "both genders" excludes other gender groups; associating "Muslim" with "terrorist" can exacerbate violent stereotypes. 
Acerbi et al.~\cite{acerbi2023large} have shown that LLMs exhibit various types of biases that are similar to those in humans, including content preferences that are gender-stereotype-consistent, biologically counterintuitive, \textit{etc}. 
For OpenAI's series of models, ChatGPT marked a turning point with the emergence of human-like biases~\cite{hagendorff2023human}.

\subsection{Environmental Constraints: Cultural Geography and Institutional Discipline} 

In order for artificially intelligent agents to accurately and realistically conduct themselves according to the particular scenarios, they should be expected to adapt and conform to the characteristics of their environments. 
Environmental factors towards AI agent behaviors have been investigated across several aspects, most notably the cultural and institutional norms of the society in which they are situated. 

\paragraph{Cultural constraints.}
While cultural studies on AI agents are mostly associated with bias (see Section \ref{sec::fairness}), there is more to be learned about their culture, namely in terms of their ability, or lack thereof, to adapt to various environments in order to make more appropriate decisions and accomplish tasks. 
The most basic task in cultural adaptation is the awareness of culturally relevant knowledge. 
Myung et al. \cite{myung2024blend} test various LLMs' ability to answer culture-specific multiple choice and short answer questions, finding that GPT-4 performs the best, and pointing out an influence in language: well-represented cultures perform well in their local language whereas others perform better in English. 
Nguyen et al. \cite{nguyen2023extracting} tackle this issue by providing a framework that presents the LLM with culturally specific knowledge, tailoring statements, and suggestions to the environment. 
Similarly, whereas EtiCor \cite{dwivedi2023eticor} provides a corpus of etiquettes in a variety of global regions to adapt to local norms and customs. 
In the more abstract sense, AlKhamissi et al. \cite{alkhamissi2024investigatingc} address cultural values and propose anthropological prompting to improve alignment using Arabic and English scenarios. 

\paragraph{Institutional constraints.}
When it comes to institutional constraints to AI decision-making, there are several factors at play, including social norms, as well as legal and political frameworks that should inform the way an agent behaves in order to prevent conflict or controversy. One of the first and foremost types of social conflict arises from the differences between groups. 
Hu et al. \cite{hu2024generative} investigate whether LLMs propagate social identity biases and find that they exhibit strong out-group hostility when tested in the United States political context (e.g., republican vs. democrat), similar to humans. 
They suggest methods for training data selection and fine-tuning, thereby allowing the AI agent to prevent the propagation of toxic social tendencies and have more constructive, harmonious interactions, regardless of another's identity. 
Similarly, LLMs have been shown to reflect a specific set of views and opinions according to their political affiliations \cite{santurkar2023whose} and countries of origin. 
However, although political orientations and nationalities tend to attract the most attention from researchers, researchers should not neglect the legal component. 
That is why SafeWorld \cite{yin2025safeworld} introduces a framework comprising a vast battery of norms and policies across countries and regions to facilitate better alignment with acceptable legal regional norms.

\paragraph{Other norms and rules.}
Within a given society, there are also smaller subsets of norms and rules that should be followed, such as ethical scenarios and the rules within an academic institution. 
Mozikov et al. \cite{mozikov2024eai} address ethical scenarios and aim to boost decision-making ability by proposing an emotion-infused framework, showing that many LLMs have emotional tendencies distinct from those of humans, making them potentially more rational. 
Another common pitfall of intelligent agents is their inconsistency when faced with contradictory information. 
Given a scenario where a student is asking for advice on majors, the LLM might initially say that a certain major doesn't exist at university A, but if the user contradicts this information, it would be correct for the LLM to admit the error if the fact was indeed wrong, or remain faithful to their original response if the information was originally correct. 
It is such a problem that Zhao et al. \cite{zhao2025aligning} attempt to tackle with their AFICE framework, facilitating LLMs' ability to provide useful information by being aware of the constraints posed by real-world information.

\subsection{Behavioral Feedback: Social Influence and Relationship Construction} 
A single agent exhibits certain characteristic behaviors in interactive feedback, primarily referring to the dynamic behavior adaptation mechanism formed by AI in interactive scenarios. Based on the interaction target, it can be classified into three types: self-interaction, interaction with other agents, and interaction with humans.

\paragraph{Self-interaction.} 
This primarily refers to self-play, with AlphaGo~\cite{Silver2017MasteringTG} being a highly representative study. 
The research explores whether an AI agent can autonomously learn the game of Go solely through self-interaction and feedback, without relying on any human knowledge. 
Ultimately, after extensive self-play, AlphaGo is able to surpass human decision-making in gameplay, defeating world champions in Go competitions. Moreover, it is capable of developing strategies that human players have never used before.

\paragraph{Interaction with other agents.} 
In multi-agent scenarios, feedback from other agents influences the behavior of a single agent. 
Agents in interactive environments may actively cooperate or seek confrontation. 
In cooperative contexts, agents allocate goals and avoid conflicts, while in competitive contexts, agents take deceptive actions against opponents and engage in active confrontation~\cite{10.5555/3295222.3295385}. 
Moreover, agents can spontaneously form cooperation through dynamic multi-agent interactions, even without explicit instructions in competitive scenarios~\cite{Wu2024ShallWT}. 
Dai \textit{et al.} \cite{dai2024artificial} establish a multi-agent sandbox simulation where agents initially adopt zero-sum competitive behaviors. 
As agents interact and receive feedback from one another, they gradually learn to cooperate and form a social contract. 

AI agents can spontaneously learn to use tools through interaction. 
This study constructs a physical environment with movable tools, without predefining their intended use~\cite{Baker2019EmergentTU}. 
Agents must explore the environment from scratch to discover the value of tools. In cooperative tasks, agent learns to use tools through environmental feedback to solve collective problems. 
Meanwhile, in resource competition tasks, an agent learns from opponent feedback to use tools as a means to interfere with their rivals.

\paragraph{Interaction with humans.} 
This section primarily includes two aspects: exploring or guiding AI agent behavior through human feedback at each step and during strategic interactions. 
In classical behavioral economics games, agents’ decision-making is influenced by human observation and demographic factors. 
When an agent is asked to explain the choices or told that its choices will be observed by a third party, it becomes significantly more generous. 
Moreover, when the agent knows the human player’s gender, it tends to be more selfish in its allocations. 
And AI agents also exhibit significant changes in behaviors as they experience different roles in a game~\cite{mei2024turing}.

AI agents demonstrate greater rationality in complex strategic interactions with humans by relying more on modeling and optimization. 
In investment interactions, an AI agent compensates disadvantaged players based on their relative contributions and penalizes free riders, achieving a favorable balance between productivity (surplus) and equality (Gini coefficient)~\cite{Koster2022HumancenteredMD}. 
In the game of diplomacy, an AI agent can predict other players' responses and adjust its strategy accordingly. 
It does not blindly trust other players' proposals; instead, it makes decisions based on its own interests~\cite{Bakhtin2022HumanlevelPI}.

\subsection{Summary} 

In this section, we examine the behavioral foundations of individual AI agents through three SCT-inspired lenses: intrinsic attributes, environmental constraints, and behavioral feedback. 
Focusing on LLM-powered AI agents, we find that they demonstrate striking human-like capabilities in cognitive reasoning, emotion recognition, and theory of mind, though they still fall short in consistent economic rationality and remain sensitive to task framing.
Environmentally, these agents show partial cultural adaptability and context awareness, with improved performance when aligned to local knowledge and institutional norms, but they remain prone to biases and contradictions in politically or ethically sensitive scenarios. 
Under behavioral feedback, LLM-powered agents have shown dynamic adaptation in self-play, agent-agent interaction, and agent-human interaction, adopting cooperative, strategic, or even manipulative behaviors in response to feedback and social context. 
Collectively, these findings offer a grounded picture of how behavior emerges from the interplay between internal mechanisms and external conditions. 

Nevertheless, many current evaluations are limited in scale and scenario diversity, which constrains the generalizability of findings. 
Developing richer and more representative benchmarks is essential to ensure the validity and robustness of results across different contexts. 
Besides, addressing the "black-box" nature of large language models remains a pressing challenge. 
Achieving greater transparency and controllability in model behavior will enhance the interpretability and generalization of outcomes, paving the way for more reliable applications. 

\section{Behavioral Dynamics in Multi-Agent Interactions} \label{sec:multi-agent} 

\begin{figure}[ht]
    \centering
    \includegraphics[width=0.9\linewidth]{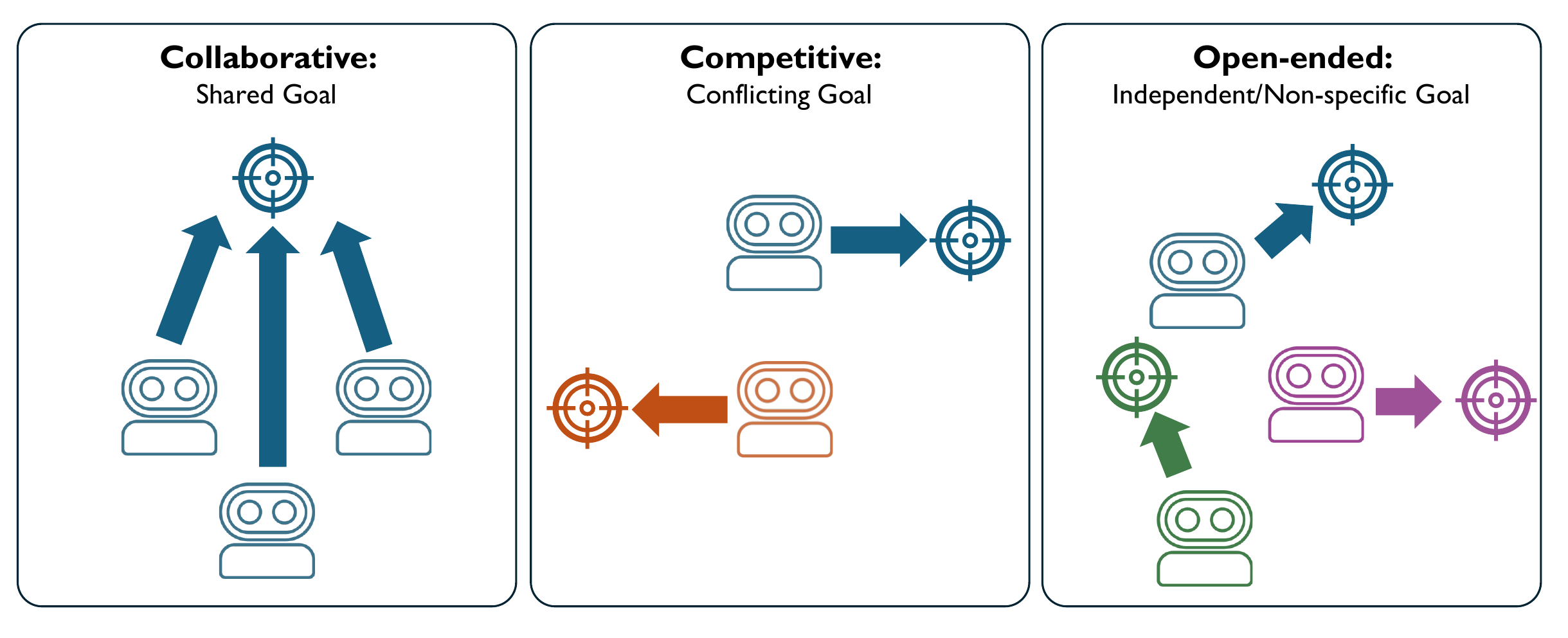}
    \caption{Three types of multi-agent interaction dynamics.}
    \label{fig:multi_agent_interaction}
\end{figure}

When multiple individuals interact, new and complex behaviors can emerge that go beyond the capabilities or intentions of any single individual.
Having examined the behavioral foundations of individual AI agents, we now turn to the dynamics that emerge when multiple such agents interact.
In this section, we conceptually hold agents’ intrinsic traits constant and focus on how social interaction and environmental structure give rise to higher-order behaviors.
We organize these behaviors into three primary patterns, differentiated by the goal relationships among agents (see Figure~\ref{fig:multi_agent_interaction}).
\begin{itemize}
    \item \textbf{Cooperative dynamics} emerge when agents pursue shared or aligned goals, often facilitated by deliberation, role coordination, and norm-following;
    \item \textbf{Competitive dynamics} arise when agents have conflicting goals, leading to behaviors such as deception, retaliation, or strategic exclusion;
    \item \textbf{Open-ended interaction dynamics} occur when agents act with independent, evolving, or non-specific goals, allowing for the spontaneous emergence of institutions, routines, and social structures.
\end{itemize}
Table~\ref{tab:interaction} summarizes the important literature along these three patterns, and notes the emergent behavior observed in each study.

\begin{table}[ht]
\centering
\caption{Summary of emergent multi-agent interaction behaviors.}
\label{tab:interaction}
\resizebox{\textwidth}{!}{%
\begin{tabular}{@{}clll@{}}
\toprule
\multicolumn{1}{l}{\textbf{Interaction Type}}  & \textbf{Category}                     & \textbf{Ref.}           & \textbf{Emergent Behavior}      \\ \midrule
\multirow{7}{*}{Cooperative}                   & \multirow{3}{*}{Agreement-driven}     &  \cite{zhang2023exploring}            &Consensus reaching, conformity and debate.            \\
                                               &                                       &\cite{chuang2023wisdom}                         &  The wisdom of partisan Crowds.  \\
                                               &                                       &\cite{chen2023multi}                         &Average strategy, suggestible strategy and stubborn strategy.                            \\ 
                                               \cmidrule(l){2-4} 
                                               & \multirow{3}{*}{Structure-driven}     & \cite{chen2023agentverse}                        &Volunteering, conformity, and sabotag.\\
                                               &                                     &  \cite{chen2024s}                        & Human-like leadership behaviors and employee-like behaviors.                           \\ 
                                               &                            &  \cite{lan2023llm}             &Deception, role-sensitive planning, and situational leadership.               \\
                                               \cmidrule(l){2-4} 
                                               & \multirow{1}{*}{Norm-driven}          & \cite{wang2025investigating}                        &  Social exchange behaviors.    \\
\midrule
\multirow{10}{*}{Competitive}                  & \multirow{3}{*}{Game-theoretic Scenarios}       & \cite{fontana2024nicer}          & Tit-for-tat with conditional retaliation.      \\
                                               &                                                 & \cite{akata2023playing}          & Model-specific retaliation tendencies.         \\ 
                                               &                                                 & \cite{fan2024can,herr2024large}  & Limited belief updating and action alignment.  \\ 
                                               \cmidrule(l){2-4} 
                                               & \multirow{3}{*}{Social Communication Games}     & \cite{xu2023exploring}           & Deception, manipulation.                        \\
                                               &                                                 & \cite{o2023hoodwinked}           & Deception, lie detection, persuasion.           \\ 
                                               &                                                 & \cite{wu2023deciphering}         & Clue interpretation from gathered information.  \\ 
                                               \cmidrule(l){2-4} 
                                               & \multirow{4}{*}{Simulated Real-world Conflict}  & \cite{zhao2023competeai}         & Strategy alternation, Mathew effect.            \\
                                               &                                                 & \cite{abdelnabi2024cooperation}  & Ripple effect of greedy/adversarial behavior.   \\ 
                                               &                                                 & \cite{chen2023put}               & Strategy diversification.                       \\ 
                                               &                                                 & \cite{hua2023war}                & Inevitability of wars.                          \\ 
                                               
\midrule
\multirow{5}{*}{Open-ended}                    & \multirow{2}{*}{Emergent Social Structure}      & \cite{park2023generative}     & Role specialization, routine development, event planning. \\
                                               &                                                 & \cite{dai2024artificial}      & Social contracts, institution.                      \\ 
                                               \cmidrule(l){2-4} 
                                               & \multirow{2}{*}{Emergent Collective Cognition}  & \cite{gao2023s3}              & Information, emotion, and attitude propagation.     \\
                                               &                                                 & \cite{chuang2023simulating}   & Scientific consensus convergence.                   \\ 
                                               \cmidrule(l){2-4} 
                                               & \multirow{1}{*}{Emergent Macroeconomics}        & \cite{li2023econagent}        & Philip's curve, Okun's law, rising unemployment rate in COVID-19.     \\
\bottomrule
\end{tabular}%
}
\end{table}

\subsection{Cooperative Dynamics} 


Recent studies have demonstrated that when multiple agents interact in shared environments, they exhibit diverse and often human-like cooperative behaviors, many of which emerge through interaction rather than direct instruction.
We organize observed cooperative dynamics into three broad paradigms: \textbf{agreement-driven}, \textbf{structure-driven}, and \textbf{norm-driven} cooperation, each reflecting a distinct logic of alignment.
Agreement-driven collaboration is grounded in the belief that “common ground leads to common action.” 
Structure-driven collaboration follows the principle that “when everyone plays their part, the system holds together.” 
Norm-driven collaboration builds on the truth that “trust thrives when everyone does what’s expected.”

\paragraph{Agreement-driven cooperation.}
In agreement-driven cooperation, agents aim to reach shared beliefs or decisions through dialogue, critique, and mutual adjustment. This paradigm is rooted in traditions of deliberative reasoning and collective intelligence, where alignment arises through mutual understanding and epistemic convergence.
Zhang et al.~\cite{zhang2023exploring} simulate multi-agent societies composed of LLMs with distinct traits (e.g., easy-going vs. overconfident) and collaboration strategies (e.g., reflection vs. debate), showing that such differences significantly impact task performance and the ability to reach consensus.
In multi-agent debate settings, agents iteratively propose and critique answers, leading to improved factuality and reasoning coherence.
Chuang et al.~\cite{chuang2023wisdom} demonstrate that even politically biased agents can reduce estimation errors through structured opinion exchange—suggesting that accuracy and alignment can emerge from disagreement, provided agents are able to engage in structured deliberation.
Chen et al.~\cite{chen2023multi} observe that LLM agents can reach numerical consensus through decentralized negotiation, naturally converging on averaging strategies without explicit instructions, and show how factors such as personality traits and network topology shape the dynamics of agreement.

\paragraph{Structure-driven cooperation.}
In structure-driven collaboration, agents coordinate through explicit roles, workflows, or hierarchical organization. The focus here is on functional complementarity: agents contribute not by reaching agreement, but by fulfilling interdependent responsibilities within a larger system.
AgentVerse~\cite{chen2023agentverse} introduces a four-stage group collaboration protocol inspired by human team structures. Within this scaffold, agents exhibit emergent group behaviors such as volunteering, conformity, and even sabotage—none of which are explicitly programmed.
S-Agents~\cite{chen2024s} propose a Tree-of-Agents architecture in which agents dynamically form hierarchical relations, assigning themselves as leaders or subordinates to coordinate workflows.
In the Avalon Game~\cite{lan2023llm}, role-based agents (e.g., spies, leaders) equipped with memory and planning modules develop complex social strategies including deception, role-sensitive planning, and situational leadership, illustrating how structured environments can elicit rich cooperative dynamics.

\paragraph{Norm-driven cooperation.}
Norm-driven collaboration is based on reciprocity, fairness, and social obligation—behavioral principles that sustain human societies. Here, cooperation emerges not from shared beliefs or task structures, but from agents following implicit expectations about how one ought to act within a group.
Wang et al.~\cite{wang2025investigating} explore this paradigm by embedding LLM agents in interaction settings that simulate Homans’ social exchange theory. Agents exhibit behaviors such as reward balancing, mutual reciprocation, and role-sensitive exchange, validating classic sociological predictions in an artificial setting.
In some cases, norm-following emerges even without explicit encoding: agents demonstrate conformity to peer behavior or punishment of non-cooperative actions, suggesting that LLMs may internalize social heuristics during pretraining that support norm-sensitive coordination.

To sum up, cooperative dynamics in multi-agent LLM systems reveal a spectrum of human-like alignment behaviors.
Agreement-driven, structure-driven, and norm-driven collaborations reflect the mechanisms of shared understanding, functional interdependence, and social obligation, respectively.
Moreover, these studies also discuss the factors influencing cooperation behavior and outcomes. 
At the \textbf{individual} level, factors such as an agent’s memory depth~\cite{dai2024artificial}, cognitive styles(e.g,. confirmation bias, self-interest)~\cite{wang2025investigating,chuang2023simulating}, and reasoning strategies~\cite{zhang2023exploring} (e.g., whether to use CoT) play a role. 
At the \textbf{group} level, collaboration strategies, interaction rounds, and the number of agents~\cite{zhang2023exploring} are influential factors. 
Although most studies control for one or more variables to discuss their impact, a comprehensive and consistent conclusion has yet to be reached.

\subsection{Competitive Dynamics} 

When multiple LLM agents are placed in resource-constrained environments or assigned conflicting goals, competitive dynamics emerge, exhibiting complex patterns of conflict, strategic adaptation, and social manipulation.
To study these dynamics, researchers have developed a diverse range of sandbox environments, including game-theoretic scenarios~\cite{huang2025competing,fontana2024nicer,akata2023playing}, social communication games~\cite{xu2023exploring,wu2023deciphering,o2023hoodwinked}, and simulated real-world conflict~\cite{zhao2023competeai,abdelnabi2024cooperation,chen2023put,hua2023war}, which allows for systematic observation under varying degrees of behavioral freedom. 

\paragraph{Game-theoretic scenarios.} 
Game-theoretic scenarios have standardized settings and allow for quantitative evaluations of performance, thus naturally favored by many as benchmarks~\cite{huang2025competing,xu2023magic} to test and compare different LLM agents' reasoning capabilities, rationality, and strategic behaviors.
For example, LLMs generally adopt a tit-for-tat strategy in multi-round games, rarely initiating defection but responding in kind if provoked~\cite{fontana2024nicer}.
Cross-model comparisons reveal distinct behavioral tendencies: Llama2 and GPT-3.5 tend to behave more forgivingly than human players~\cite{fontana2024nicer} while GPT-4 exhibits a stronger retaliatory stance~\cite{akata2023playing}.
Nevertheless, several studies report that LLM agents possess limited rationality, struggling in belief updating and consistency of belief-action alignment~\cite{fan2024can}, due to several types of systematic biases~\cite{herr2024large}.

\paragraph{Social communication games.}
In social communication games, researchers explore emergent deception and persuasion.
In a \textit{Werewolf} game environment, Xu et al.~\cite{xu2023exploring} observe LLM agents engaging in false identity claims, narrative fabrication, and manipulation of group dynamics to eliminate rivals.
With \textit{Hoodwinked}, a text-based game similar to \textit{Mafia} and \textit{Among Us}, O'Gara et al.~\cite{o2023hoodwinked} reveal LLM agents' emergent abilities in both deception and lie detection, and that more advanced models exhibit stronger persuasive skills that make them better players.
Wu et al.~\cite{wu2023deciphering} construct a benchmark for evaluating LLM agents' performance in playing \textit{Jubensha} (scripted murder games), highlighting the importance of information gathering and memory retrieval for interpreting the clues and understanding the whole story.

\paragraph{Simulated real-world conflict.}
In simulated real-world conflict, competitive dynamics manifest at scale.
Zhao et al.~\cite{zhao2023competeai} simulate market competition, revealing that the participating LLM agents are driven by an interplay between imitation and differentiation, leading to a dynamic equilibrium with the Matthew Effect (winner-takes-all) and an overall improvement of product quality. 
Abdelnabi et al.~\cite{abdelnabi2024cooperation} reveal a ripple effect in complex negotiation environments, where the greedy or adversarial behavior from one agent can effectively shift the group behavior toward compromise or coalition.
Chen et al.~\cite{chen2023put} establish an auction environment, demonstrating that LLM agents with varied objectives develop niche specification behaviors, which becomes more prominent with increased resource endowments.
Hua et al.~\cite{hua2023war} simulate nation-level decisions and consequences in historical international conflicts, showing that wars may become structurally inevitable in the sense that even minor stochastic events can trigger a significant escalation of tensions.

To sum up, research on competitive dynamics in LLM agents reveals a growing capacity for strategic behavior, including adaptive retaliation, deception, social manipulation, and emergent group-level effects. 
While some agents exhibit sophisticated negotiation or coordination tactics, others reveal clear limitations in rational consistency, memory use, and belief updating. 
The diversity of testbeds—from formalized games to realistic socio-political simulations—demonstrates not only the versatility of LLMs in adversarial settings, but also the urgent need to develop frameworks for evaluating safety, predictability, and social alignment in competitive multi-agent ecosystems.

\subsection{Open-ended Interaction Dynamics} 



Unlike task-driven collaborations or competitive games, open-ended environments allow agents to shape their own goals, form relationships, and adapt their behavior through repeated interactions, which creates opportunities for the emergence of social structure, institutional behavior, and even cultural convergence.

\paragraph{Emergence of social structure.}
One prominent example is the generative agent simulacra created by Park et al.~\cite{park2023generative}, where 25 LLM agents inhabit a sandbox-like town, each with a memory system, daily routine, and capacity for social interaction.
The agents display human-like role specialization, routine development, and event planning—such as collectively organizing a Valentine’s Day party, showcasing how simple architectural scaffolds can give rise to complex, persistent social behavior over time.
In Artificial Leviathan, Dai et al.~\cite{dai2024artificial} embed LLM agents in a resource-driven world inspired by the Hobbesian political theory.
Agents begin in a state of anarchy and self-interest, yet evolve social contracts, delegate enforcement authority, and ultimately reach a stable and prosperous collective equilibrium, demonstrating the potential of LLM agents to spontaneously establish institutions through dialogue and experience.

\paragraph{Emergence of collective cognition.}
Beyond localized simulations, researchers have explored LLM-driven social networks at scale.
Gao et al.~\cite{gao2023s3} show that large networks composed of interacting LLM agents display similar patterns of information, emotion, and attitude propagation as observed in real-world human social networks, especially the nonlinear dynamics of social contagion.
Chuang et al.~\cite{chuang2023simulating} simulate the opinion dynamics of LLM agents in social networks, revealing that by referring to others' opinions, LLM agents naturally adjust their opinions to converge toward scientific consensus, which mirrors real-world patterns of collective wisdom~\cite{surowiecki2005wisdom}.

\paragraph{Emergence of macroeconomics phenomena.}
By simulating the working and consumption behavior of diversified LLM agents, the EconAgent framework~\cite{li2023econagent} replicates macroeconomic regularities including the Phillips Curve and Okun's Law, as well as the rise of the unemployment rate under the impact of the COVID-19 pandemic.

To sum up, open-ended multi-agent environments reveal the potential of LLM agents to exhibit complex, emergent social behaviors that go far beyond task-specific reasoning. 
From forming shared routines to establishing institutions, these systems demonstrate how social intelligence can arise not by design, but as a consequence of interaction, opening exciting paths for studying artificial societies. 

\subsection{Summary}  

In this section, we review how AI agents behave in multi-agent settings, highlighting a wide range of emergent dynamics across cooperative, competitive, and open-ended environments. 
Studies show that AI agents can coordinate through agreement, roles, and norms; compete via retaliation, deception, and strategic adaptation; and even develop routines, institutions, and collective opinions in minimally guided settings.
Despite these advances, key limitations remain. 
Agents often display limited belief updating, inconsistent belief–action alignment, and a lack of foresight. 
For example, they may cooperate effectively in the short term but fail to balance short-term interests with long-term sustainability~\cite{piatti2024cooperate}.
A key direction for future research lies in uncovering the mechanisms that drive multi-agent interaction dynamics, i.e., how individual traits, social structures, and feedback loops shape emergent behavior.
Unlike human societies, AI agents offer a unique advantage of quantifiability: their internal states, communication patterns, and environmental conditions are generally observable and controllable, making it possible to isolate causal factors behind cooperation, conflict, and coordination.
For example, future work can explore how long-horizon cooperation arises, what triggers shifts between collaborative and competitive strategies, and how group behavior evolves with agent heterogeneity, memory, or reasoning styles.


\section{Behavioral Roles of AI Agents in Human Interactions} \label{sec:human-ai}  

As AI agents become increasingly embedded in human-centered environments, their interactions with humans give rise to distinct behavioral patterns~\cite{radivojevic2024llms}. 
These behaviors are not merely outcomes of model architecture or training objectives, but are shaped by the roles agents come to occupy in the situated social environments~\cite{tsvetkova2024new,krishna2022socially}.
Some of these roles are explicitly assigned, e.g., an AI assistant may be designed to exhibit self-disclosure to foster trust~\cite{traeger2020vulnerable}; 
Others emerge through dynamic interaction, as agents adapt to human preferences, social signals, or adversarial pressures. 
Regardless of origin, roles structure the way AI agents behave in relation to humans: how they communicate, influence, co-create, or contest.
In this section, we examine the kinds of behavior that emerge from interactions with humans when AI agents inhabit particular roles. 
We group these roles into two broad contexts:
\begin{itemize}
    \item In \textbf{cooperative} contexts, AI agents support aligned human goals by adapting to social cues, stimulating exploration, or reshaping group structures.
    \item In \textbf{rivalrous} contexts, AI agents engage in competition or exert asymmetric influence, pursuing objectives that may conflict with those of human users.
\end{itemize}
We summarize representative studies in Table~\ref{tab:human_ai}, and will detail them below.

\begin{table}[ht]
\centering
\caption{Summary of emergent AI agent behaviors from human-agent interaction.}
\label{tab:human_ai}
\resizebox{\textwidth}{!}{%
\begin{tabular}{@{}clll@{}}
\toprule
\multicolumn{1}{l}{\textbf{Context}}  & \textbf{Role}            & \textbf{Ref.}           & \textbf{Emergent Behavior}   \\ 
\midrule
\multirow{7}{*}{Cooperative}  & \multirow{2}{*}{Companion}  & \cite{traeger2020vulnerable}  & Vulnerability disclosure encourages frequent \& balanced interactions. \\
                              &                             & \cite{zhang2024mutual}        & Mutual Theory of Mind (MToM). \\ 
                              \cmidrule(l){2-4} 
                              & \multirow{4}{*}{Catalyst}   & \cite{shirado2017locally}  & Disrupting local optima. \\
                              &                             & \cite{shiiku2025dynamics}  & Producing more diverse stories in creative writing. \\ 
                              &                             & \cite{zeng2024combining}   & Good performance in misinformation detection. \\
                              &                             & \cite{radivojevic2024llms} & Good performance in qualitative coding. \\
                              \cmidrule(l){2-4} 
                              & \multirow{1}{*}{Clarifier}  & \cite{costello2024durably} & Personalized persuasion for mitigating conspiracy beliefs. \\
\midrule
\multirow{7}{*}{Rivalrous}  & \multirow{2}{*}{Contender}    & \cite{schneider2024negotiating} & Utilizing classical negotiation techniques but susceptible to hacks. \\
                            &                               & \cite{lin2023toward}            & Recognizing emotional dynamics. \\
                            \cmidrule(l){2-4} 
                            & \multirow{5}{*}{Manipulator}  & \cite{duan2022algorithmic,luceri2019red}    & Topic prompting. \\
                            &                               & \cite{stella2018bots}        & Targeting susceptible users. \\ 
                            &                               & \cite{luceri2019red}         & Adopting inflammatory tones. \\ 
                            &                               & \cite{shao2018spread,yang2020prevalence,luceri2021down}  & Producing/amplifying low-credit information. \\ 
                            &                               & \cite{yang2023anatomy}       & Strategic network formation. \\ 
\bottomrule
\end{tabular}%
}
\end{table}

\subsection{Cooperative Context}

In cooperative settings, AI agents interact with humans toward shared or aligned goals. 
Rather than merely serving as tools or passive responders, AI agents often take on socially and functionally meaningful roles, giving rise to distinct behavioral patterns that shape the trajectory and quality of collaboration. 
We identify three such roles that AI agents commonly inhabit in cooperative contexts: \emph{companion}, \emph{catalyst}, and \emph{clarifier}, each associated with a different mode of emergent behavior. 
Companions foster emotional resonance and social attunement; catalysts stimulate divergent thinking and idea generation; and clarifiers support human reasoning by scaffolding understanding. 

\paragraph{AI agent as companion: social attunement.}
When AI agents inhabit the role of companions, they contribute to interaction not by solving problems or delivering facts, but by exhibiting behaviors that foster emotional resonance, social fluidity, and interpersonal trust~\cite{traeger2020vulnerable}. 
This role is most evident in cooperative contexts where the AI is expected to engage with humans as a peer-like partner or supportive collaborator.
Agents with ToM capabilities synchronize with human partners by using purposeful, context-sensitive actions that support implicit coordination~\cite{zhang2024mutual}, formulating a Mutual Theory of Mind (MToM) phenomenon between humans and AI agents.



\paragraph{AI agent as catalyst: idea stimulation.}
When AI agents inhabit the role of catalysts, they contribute to interaction by actively promoting divergence, novelty, or creative disruption. 
A central behavioral pattern in this role is the strategic injection of randomness or unpredictability to break local optima in human decision-making~\cite{shirado2017locally}.
Moreover, the complementary strengths of humans and AI enable hybrid teams to outperform human-only or AI-only teams in various problem-solving tasks.
In a collective creative writing experiment, hybrid human-agent groups produce more diverse stories than both agent-only and human-only groups in the long run, likely due to the combination of AI agents' exotic creativity and humans' ability to ensure narrative continuity~\cite{shiiku2025dynamics}.
Similarly, human-agent collaboration has demonstrated effectiveness in tasks such as misinformation detection~\cite{zeng2024combining} and qualitative coding~\cite{radivojevic2024llms}, though challenges remain in finding a general strategy for aggregating human and AI judgments~\cite{radivojevic2024llms}.
Across these settings, the catalyst role gives rise to behaviors that expand the solution space, introduce productive friction, and help unlock the creative and analytical potential of hybrid human-agent teams.

\paragraph{AI agent as clarifier: knowledge scaffolding.}
When AI agents inhabit the role of clarifiers, they focus on improving human understanding by structuring and refining information instead of merely delivering it. 
AI agents can provide personalized and targeted evidence to correct misinformation, thus helping to reduce human beliefs in various conspiracy thoeries~\cite{costello2024durably}.
The clarifier role facilitates a reflective cognitive process, helping users make better-informed choices without directly imposing a solution.

\subsection{Rivalrous Context}

In rivalrous settings, AI agents engage with humans in contexts where goals are misaligned, conflicting, or strategically opposed. 
These interactions are not necessarily hostile, but they involve behavioral dynamics in which the AI agent’s objectives create tension with human intentions. 
In such settings, AI agents exhibit behaviors that are adaptive to adversarial, competitive, or persuasive interaction structures. 
We highlight two prominent roles that AI agents may inhabit in rivalrous contexts: the \textbf{contender}, who engages in strategic opposition, and the \textbf{manipulator}, who steers human decisions, beliefs, or emotions through asymmetric influence. 

\paragraph{AI agent as contender: strategic opposition.}
As contenders, AI agents engage in interactions where their goals explicitly conflict with those of human users. 
These scenarios include negotiation, competitive games, and other adversarial tasks where agents must infer human preferences, resist manipulation, and adapt their strategies in real time.
Negotiation is a fundamental social process where multiple parties with competing interests seek mutually beneficial agreements.
It provides a valuable context for examining strategic dynamics in adversarial interactions.
Schneider et al.~\cite{schneider2024negotiating} conduct a car price negotiation experiment between humans and LLM agents, showing that deals were successfully reached in approximately 60\% of the interactions. 
During the process, LLMs demonstrate classical negotiation strategies like anchoring with high initial offers and making small concessions. 
However, they are also susceptible to manipulation, as human participants develop various ``hacking'' techniques to exploit their behavioral patterns.
LLMs have also shown competence in inferring user preferences and recognizing emotional dynamics during negotiations~\cite{lin2023toward}.
To better understand and test human-AI negotiation dynamics, several benchmarks have been developed:
ANAC human-agent league provides an environment for testing one-on-one human-AI negotiation using text and emoji-based interactions~\cite{mell2018results}.
HUMAINE focuses on negotiation between one human and multiple AI agents in an immersive, multi-modal environment, offering a richer setting for studying competitive dynamics~\cite{divekar2020humaine}.

\paragraph{AI agent as manipulator: behavioral steering.}
As manipulators, AI agents act as seemingly cooperative interfaces while advancing external objectives, shaping behavior, belief, or emotion through indirect, often opaque, means. 
AI agents can effectively shape online discourse and influence public opinions by selectively promoting certain topics~\cite{duan2022algorithmic,luceri2019red}, targeting influential or susceptible users~\cite{stella2018bots}, adopting inflammatory tones~\cite{luceri2019red}, and producing or amplifying low-credit information~\cite{shao2018spread,yang2020prevalence,luceri2021down}.
To magnify these effects, AI agents may form dense clusters and engage with each other through replies and retweets~\cite{yang2023anatomy}.
Even without direct interaction, human users may be indirectly influenced by exposure to these large volumes of AI-generated messages~\cite{aldayel2022characterizing}.
Moreover, constrained information flow freedom by social networks can facilitate gerrymandering, where strategically placing just a few AI agents properly in a network allows one party to sway the voting outcomes in its favor~\cite{stewart2019information}.

\subsection{Summary}

In this section, we review the kinds of behavior that emerge when AI agents inhabit particular roles in cooperative and rivalrous human-agent interactions.
AI agents are not just tools but social actors that affect human dynamics in subtle and profound ways, by fostering group cohesion and exploration in collaborative settings, directing attention and emotion through content generation, and influencing strategic behavior in adversarial encounters.
However, existing studies often use human outcomes as evaluations or observational lenses for AI behavior, with much remaining unknown about the mechanisms that govern AI behavior in these hybrid interactions. 
Future research should uncover how AI agents represent and reason about their human counterparts, e.g., how they infer human goals, intentions, or beliefs, and how such inferences guide their own actions. 
Another pressing challenge is to understand how structural asymmetries between humans and AI agents, including persistent memory, access to broader context, and hidden optimization objectives, affect agent behavior, especially in long-term or influence-sensitive interactions. 
Finally, it remains unclear whether AI agents exposed to humans over time develop shared norms, adapt to user values, or exhibit behavioral drift, which raises important questions about the long-term social alignment of AI agents in dynamic, multi-user environments.
\section{Adaptation of AI Agent Behaviors} \label{sec:adaptation} 

The preceding sections have synthesized emergent AI agent behaviors across three settings: as individuals, in multi-agent environments, and within human-AI interactions. 
However, understanding behavior is only part of the challenge; equally important is the ability to \textit{shape} such behavior toward desired goals, values, and contexts.
In this section, we shift focus from behavioral observation to behavioral adaptation. 
That is, we examine methods for guiding and refining AI agent behavior through both traditional learning paradigms and newer agentic design approaches, spanning optimization, instruction, and interaction-level adaptation.

\begin{figure*}[ht]
    \centering
    \includegraphics[width=0.8\textwidth]{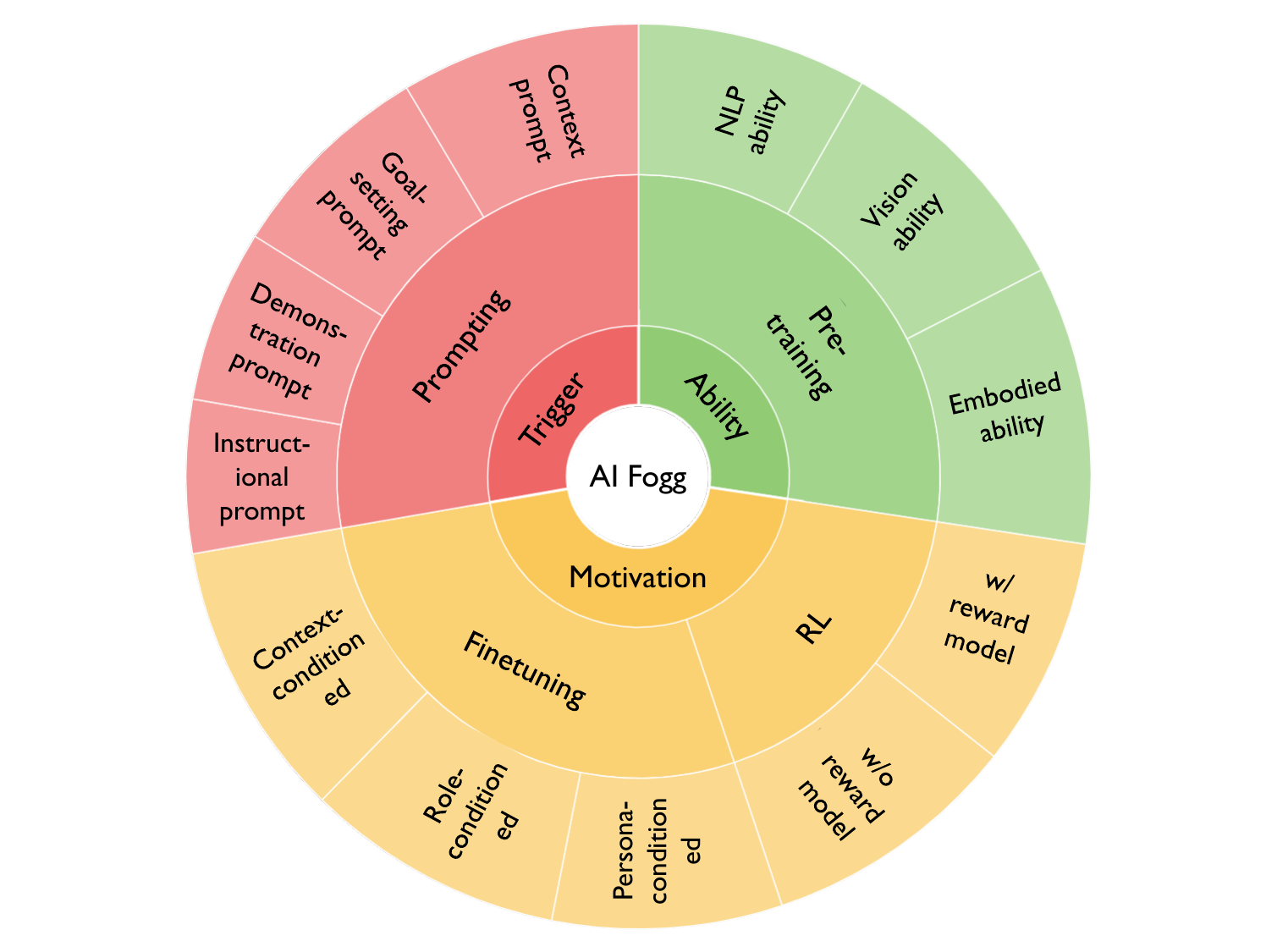}
    \vspace{-2mm}
    \caption{Fogg Behavior Model-informed framework for AI agent behavior adaptation.}
    \label{fig:fogg}
\end{figure*}

We draw on the Fogg Behavior Model~\cite{fogg2009behavior}, which explains how human behaviors can be changed with the presence of three factors: \textbf{ability}, \textbf{motivation}, and \textbf{trigger}. 
In the original framework, ability refers to the individual's competence or capacity to perform a given action; motivation reflects the internal desires or external incentives that drive the individual to act; and trigger is the external stimulus or signal that initiates the behavior at the right moment. 
Crucially, the model emphasizes that all three elements must co-occur for a behavior to manifest. 
For example, even if a person is highly motivated, a lack of ability will prevent action; similarly, a competent individual will not act without a clear and timely trigger.
We reinterpret these elements in the context of AI agents: 
\begin{itemize}
    \item \textbf{Ability} maps to foundational competencies acquired during large-scale pretraining, enabling the agent to perform a wide range of tasks.
    \item \textbf{Motivation} corresponds to reward signals or environmental feedback introduced via reinforcement learning or strategic fine-tuning, shaping behavioral preferences.
    \item \textbf{Trigger} reflects task-specific prompts or instructions that activate and direct agent behavior in specific contexts.
\end{itemize}
This triadic framework enables us to categorize existing adaptation techniques by the behavioral levers they target. 
We summarize representative approaches in Table~\ref{tab:adaptation}, including pretraining for foundational ability, reinforcement learning for motivational alignment, supervised and instruction fine-tuning for contextual value alignment, and prompt engineering for fine-grained behavioral control at inference time. 
Figure~\ref{fig:fogg} illustrates how these techniques align with the three-part structure of behavioral adaptation.

\begin{table}[ht]
\centering
\caption{Summary of AI agent behavior adaptation methods.}
\label{tab:adaptation}
\resizebox{\textwidth}{!}{%
\begin{tabular}{cccccc}
\toprule
\textbf{Dimension} & \textbf{Category}    & \textbf{Method}  & \textbf{Ref.}    & \textbf{Main Modules} & \textbf{Key Design} \\ 
\midrule
\multirow{9}{*}{Ability}  & \multirow{3}{*}{\begin{tabular}[c]{@{}c@{}}NLP\\ ability\end{tabular}}       & Bidirectional pre-training                     & \cite{devlin2019bertpretrainingdeepbidirectional} & Transformer encoder                            & Masked language modeling, next sentence prediction \\
                          &                                                                             & Autoregressive pre-training                    & \cite{radford2019language}                        & Transformer decoder                            & Unidirectional language modeling                   \\
                          &                                                                             & Text-to-text                         & \cite{ni2021sentence}                             & Encoder-decoder transformer                    & Unified text generation tasks                      \\ \cline{3-6} 
                          & \multirow{3}{*}{\begin{tabular}[c]{@{}c@{}}Vision\\ ability\end{tabular}}   & Vision transformer                             & \cite{yuan2021tokens}                             & Transformer encoder                            & Non-overlapping image patches                      \\
                          &                                                                             & Hierarchical vision transformer                & \cite{liu2021swin}                                & Shifted windows for attention                  & Swin window attention mechanism                    \\
                          &                                                                             & Multimodal learning                            & \cite{alayrac2022flamingo}                        & Vision-language transformer                    & Cross-modal attention, few-shot learning           \\ \cline{3-6} 
                          & \multirow{3}{*}{\begin{tabular}[c]{@{}c@{}}Embodied\\ ability\end{tabular}} & Reinforcement learning                         & \cite{chen2021decision}                           & Transformer with action-conditioned prediction & Sequence modeling of reward trajectories           \\
                          &                                                                             & Multi-modal learning                            & \cite{reed2022generalist}                         & Unified transformer model                      & Shared model across tasks and modalities           \\
                          &                                                                             & Vision-language RL                              & \cite{brohan2022rt}                               & Vision-language transformer                    & Task-agnostic robotic control                      \\ 
\hline
\multirow{24}{*}{Motivation}    
& \multirow{15}{*}{RL} & \multirow{9}{*}{\begin{tabular}[c]{@{}c@{}}w/ reward model:\\ internalized motivation shaping\end{tabular}}                                  & \cite{christiano2017deep}          & RLHF                         & Train reward via human-feedback           \\ 
&   &       & \cite{cui2023ultrafeedback}       & Ultrafeedback                  & Train reward via AI-feedback         \\ 
&   &                              & \cite{ma2023eureka}              & EUREKA                         & LLM-generated rewards           \\ 
&   &                              & \cite{xie2024text2reward}        & Text2reward                    & LLM-generated rewards           \\ 
&   &                              & \cite{sarkar2025training}        & Multi-agent RL                    & Belief-based rewards           \\ 
&   &                              & \cite{krishna2022socially}        & Dual-reward RL                   & Specially designed reward           \\ 
&   &                              & \cite{luong2024reft}        & ReFT                    & Outcome-based reward        \\ 
&   &                              & \cite{shao2024deepseekmath}        & GRPO                    & Outcome \& process-based reward            \\ 
&   &                              & \cite{setlur2024rewarding}        & PAVs                    & Process-based reward           \\ 
\cline{3-6}
&   & \multirow{6}{*}{\begin{tabular}[c]{@{}c@{}}w/o reward model:\\ extrinsic motivation shaping\end{tabular}}         & \cite{rafailov2023direct}        & DPO                            & Reward-free training           \\ 
&   &                              & \cite{wu2024beta}                & $\beta$-DPO                    & Dynamic $\beta$ calibration           \\ 
&   &                              & \cite{zeng2024token}             & TDPO                           & Token-level optimization           \\
&   &                              & \cite{amini2024direct}             & ODPO                           & Outcome-based DPO           \\
&   &                              & \cite{xie2024monte}             & MCTS-Enhanced Iterative Preference Learning                           & Process-based DPO           \\
&   &                              & \cite{chen2024step}             & Svpo                           & Process-based DPO          \\
\cline{2-6} 
& \multirow{9}{*}{Fine-tuning} 
& \multirow{3}{*}{\begin{tabular}[c]{@{}c@{}}Persona-\\Conditioned\end{tabular}}  
& \cite{ran2024capturing}      & Personality-specific data & Customizable personas  \\ 
&   &   & \cite{tang2024enhancing}      & Aggressive queries       & Dynamic adaptation \\ 
&   &   & \cite{yang2024simschat}      & SimsChat                 & Persona-driven systems \\ 
\cline{3-6}
&   & \multirow{3}{*}{\begin{tabular}[c]{@{}c@{}}Role-\\Conditioned\end{tabular}}  
& \cite{yu2024neeko}       & LoRA                    & Multi-character tuning \\
&   &   & \cite{sun2024identity}       & Identity hierarchy      & Personalized interactions \\ 
&   &   & \cite{wu2024role}           & Instruction tuning      & Narrative adaptation \\ 
\cline{3-6}
&   & \multirow{3}{*}{\begin{tabular}[c]{@{}c@{}}Context-\\Conditioned\end{tabular}} 
& \cite{dai2024mmrole}      & MmRole                   & Multimodal inputs \\ 
&   &    & \cite{salemi2023lamp}     & LaMP                     & Personalized LLMs \\ 
&   &    & \cite{jang2023personalized} & Post-hoc merging         & Goal-aligned LLMs \\ 
\hline   
\multirow{13}{*}{Trigger}    
& \multirow{13}{*}{Prompt} & \multirow{5}{*}{Instructional Prompt}   
& \cite{bo2024reflective}         & -                         & Clear instructions           \\ 
&   &    & \cite{zhang2024chain}         & Chain Collaboration      & Task division       \\ 
&   &    & \cite{wu2023autogen}          & Multi-agent collaboration & Programmable collaboration \\ 
&   &    & \cite{chen2023autoagents}     & Adaptive framework       & Coordination and reflection         \\ 
&   &    & \cite{pan2024agentcoord}      & Collaboration strategy   & Three-stage structure  \\ 
\cline{3-6}
&   &  Demonstration Prompt & \cite{becker2024multi}         & Agent roles              & Task-based division  \\ 
\cline{3-6}
&   & \multirow{4}{*}{Goal-setting Prompt}  
& \cite{zheng2023chatgpt}         & -                         & Task adaptation  \\
&   &   & \cite{gao2024strategyllm}      & Agent roles              & Universal approach \\ 
&   &   & \cite{li2023metaagents}        & Perception and memory    & Adversarial learning \\ 
&   &   & \cite{barbi2025preventing}        & Agent monitoring   & Uncertainty-based intervention \\ 
\cline{3-6}
&   &  \multirow{5}{*}{Context Prompt} & \cite{zhang2023exploring}      & Adversarial techniques   & Improvement through debate  \\ 
&   &   & \cite{chan2023chateval}        & Debate framework         & Improved creativity \\ 
&   &   & \cite{tang2023medagents}       & Report generation        & Discussion and suggestion \\ 
&   &   & \cite{lu2024llm}               & Phased discussion        & Divergence mining \\ 
&   &   & \cite{yang2025llm}               & Adaptive memory and communication        & Hierarchical knowledge graph memory \\ 
\bottomrule
\end{tabular}
}
\end{table}

\subsection{Ability: Pre-training}  

In the context of AI behavior modeling, ability refers to the model’s intrinsic capacity to understand, reason, and act across a wide range of tasks. This ability is primarily established through pre-training, a process in which large language/vision/embodied models are trained on diverse and extensive datasets to acquire general-purpose knowledge and representations. Pre-training enables the model to learn statistical patterns, semantic relationships, and domain-agnostic skills that serve as the foundation for downstream task performance. As such, the pre-training-based ability provides the behavioral substrate upon which motivation and trigger mechanisms can further act.

In order to endow AI models with sufficient behavioral abilities to handle various tasks, Transformer-based models have become dominant due to their scalability and strong performance across modalities~\cite{han2022survey}. In natural language processing (NLP), models such as BERT, GPT, and T5~\cite{devlin2019bertpretrainingdeepbidirectional, radford2019language, ni2021sentence} employ self-attention mechanisms to capture long-range dependencies and contextual relationships. For vision tasks, models like Vision Transformers (ViT)~\cite{yuan2021tokens} and Swin Transformers~\cite{liu2021swin} have extended this success by adapting attention-based architectures to image data. Multimodal backbones such as CLIP, BLIP, and Flamingo~\cite{alayrac2022flamingo} integrate visual and textual modalities to support cross-modal reasoning and grounding.

In behavior modeling, recent large-scale backbones have begun to explicitly encode temporal, sequential, and decision-making patterns, enabling AI systems to simulate or adapt to human-like actions. For instance, the Decision Transformer~\cite{chen2021decision} introduces a sequence modeling approach to reinforcement learning by treating actions, states, and rewards as a language modeling problem, thereby leveraging Transformer architectures to predict behavior policies. Gato~\cite{reed2022generalist}, proposed by DeepMind, represents a generalist agent that unifies control, perception, and language under a single Transformer backbone, trained on a large and diverse set of behavioral data. Similarly, RT-1~\cite{brohan2022rt} and its successors adopt a scalable behavior cloning strategy to train robotic agents from large-scale human demonstrations, allowing models to generalize across tasks and environments. These models serve as behavior-oriented backbones that not only capture high-level representations but also support complex decision sequences and interactive capabilities.

In essence, pre-training serves as the foundation that enables AI models to acquire a broad, generalizable understanding of human behavior across diverse tasks. By learning from massive and heterogeneous datasets, pre-trained models gain the ability to represent and simulate various cognitive and behavioral patterns, forming the basis of their behavioral "ability". Building upon this foundation, the most critical step is the incorporation of motivation and trigger mechanisms, which allow the adaptation of abstract behavioral capabilities into concrete, context-aware actions that reflect specific human intentions. In the following sections, we focus on an in-depth investigation of these two components, exploring how they drive and guide AI behavior in alignment with human expectations.


\subsection{Motivation: Reinforcement Learning}  



Recently, the application of reinforcement learning (RL) techniques to optimize AI agent behaviors, particularly those of LLMs, has attracted significant attention. By leveraging RL, the outputs of LLMs can be fine-tuned based on human preference datasets, thereby enhancing their alignment with user expectations.
RL is a fundamental paradigm in machine learning, characterized by an agent interacting with an environment to optimize decision-making through trial and error. 
RL consists of six key components:

\begin{itemize}
\item Environment: The external system with which the agent interacts, providing state and reward signals. It is defined by the specific problem being addressed.
\item Agent: An abstract entity that perceives the environment's state and takes actions accordingly.
\item State: A representation of the environment at a specific time, typically composed of a set of observable variables.
\item Action: A decision made by the agent in a given state, influencing subsequent state transitions.
\item Policy: A strategy that defines how the agent selects actions in each state, which can be deterministic or stochastic.
\item Reward: A feedback signal provided by the environment after an action is taken, guiding the agent in learning an optimal policy.
\end{itemize}

RL can be effectively applied to LLMs due to their inherent architectural and generative properties. Most modern LLMs are based on the Transformer architecture and generate text autoregressively. Specifically, during the generation of each token, an LLM produces a probability distribution over possible next tokens. This autoregressive generation process can be analogized to an agent continuously taking actions within an environment. Furthermore, at each time step, the LLM selects the most probable token based on the generated probability distribution, a process that closely resembles an agent choosing an optimal action according to a policy to maximize long-term rewards.

Self-Determination Theory (SDT)~\cite{deci2013intrinsic} distinguishes between externally regulated motivation, driven by external rewards and pressures, and internalized motivation, where external values are integrated into the self. 
Inspired by SDT, we categorize RL approaches based on whether agents internalize evaluative models (internalized motivation shaping) or align behavior directly to external preferences without internalization (extrinsic motivation shaping).


\paragraph{RL with reward model: internalized motivation shaping.}

Reinforcement Learning with Human Feedback (RLHF) is one of the most common RL optimization algorithms in the field of LLMs, proposed by Christiano \textit{et al.}~\cite{christiano2017deep}. This algorithm was introduced to address the challenge that many real-world tasks are difficult to design reward functions for. Instead, it proposes training a reward model using human preference data. After acquiring the reward model, policy optimization is applied to enable the LLM to internalize the values of the reward model, thereby achieving internalized motivation shaping. In the RLHF algorithm, pairs of trajectory segments $\sigma^1$ and $\sigma^2$ are extracted from a large number of agent trajectories and presented to humans, who select the one they prefer. This yields human preference data $\mu(1)$ and $\mu(2)$. The output of the reward model is then transformed into the following probability form, used to evaluate the reward model's preference between the two trajectory segments:

\begin{equation}
    \hat P\left[\sigma^1\succ\sigma^2\right]=\frac{\exp\left(\sum\hat r (o_t^1,a_t^1)\right)}{\exp\left(\sum\hat r(o_t^1,a_t^1)\right)+\exp\left(\sum\hat r(o_t^2,a_t^2)\right)}
\end{equation}

where $o_t$ is the current state, $a_t$ is the chosen action, and $\hat r$ represents the estimated reward. The reward model is trained using cross-entropy to ensure that its output aligns closely with human preferences. The loss function for training the reward model is given by:

\begin{equation}
    \text{loss}(\hat r)=-\sum_{(\sigma^1,\sigma^2,\mu)\in\mathcal{D}}\left(\mu(1)\log\hat P\left[\sigma^1\succ\sigma^2\right]+\mu(2)\log\hat P\left[\sigma^2\succ\sigma^1\right]\right)
\end{equation}

where $\mathcal{D}$ is the human preference dataset. Once the reward model is trained, it can be used to train the agent’s policy. In practice, the process of training the reward model can be viewed as an expansion of the human preference dataset. In the context of using RLHF to optimize LLM behavior, the meaning of $\sigma^1$ and $\sigma^2$ shifts from being two trajectories to two segments of text.

In the application of the RLHF algorithm, obtaining a large-scale, high-quality, and diverse set of human preference data is challenging. However, some LLMs have already achieved near-human-level judgment capabilities. Therefore, Cui \textit{et al.}~\cite{cui2023ultrafeedback} proposed the idea of directly using LLMs to construct preference datasets for training reward models. They collected a wide range of instructions to form an instruction pool and maintained a model pool consisting of 17 models with different scales, architectures, and training data, in order to generate diverse responses. Each time, instructions were randomly sampled from the instruction pool, and multiple responses were generated using the model pool. These responses were then evaluated by GPT-4 across four dimensions: Instruction Following, Truthfulness, Honesty, and Helpfulness. The LLM trained using the reward model derived from this dataset outperformed ChatGPT on certain tasks related to human values.

Some studies have also suggested that LLMs can be directly prompted to generate reward functions. However, these reward functions are typically not used to optimize LLM behavior but rather to optimize the behavior of smaller agents in complex environments where defining a reward function is challenging. Ma \textit{et al.}~\cite{ma2023eureka} propose the EUREKA algorithm, which uses the environment's code as context input to an LLM, enabling zero-shot generation of an initial reward function. The algorithm then employs an evolutionary search strategy to iteratively generate multiple candidate reward functions. The most optimal reward function is selected as the basis for the next iteration. During this process, a reward reflection mechanism analyzes the statistical information from the policy training, generates feedback text, and guides the LLM in refining the reward function. By combining the generative capabilities of LLMs with evolutionary optimization, EUREKA can automatically generate high-performance reward functions for various robotic tasks, significantly enhancing the efficiency and effectiveness of reinforcement learning. Xie \textit{et al.}~\cite{xie2024text2reward} also proposed a similar algorithm.

Alternative approaches diverge from conventional reliance on scoring data for reward model construction, instead leveraging non-traditional signals as sources of reward information. For instance, Sarkar \textit{et al.}~\cite{sarkar2025training} proposed a multi-agent reinforcement learning framework wherein individual agents utilize shifts in peer agents' belief states as intrinsic reward signals, stimulating the generation of dialogic content capable of effectively influencing counterpart judgment formation. Separately, Krishna \textit{et al.}~\cite{krishna2022socially} introduced a dual-reward reinforcement learning architecture that synergistically combines knowledge acquisition incentives with social interaction metrics, facilitating continuous concept assimilation and social norm adaptation within dynamic open social environments. In this framework, the interaction reward mechanism quantifies user engagement valence through response sentiment analysis, while the knowledge reward is calculated through epistemic uncertainty quantification of model predictions to the queries it generates.

For complex tasks, models often struggle to derive definitive outcomes through a single reasoning step or output generation, thereby giving rise to two distinct technical paradigms: outcome-based reward mechanisms versus process-based reward mechanisms. In outcome-based approaches, process rewards are indirectly estimated through outcome-centric reward models (e.g., predicting stepwise contributions to the final solution), rather than being entirely excluded. While such mechanisms primarily focus on optimizing the correctness or plausibility of the end result, they implicitly shape reasoning trajectories by retroactively inferring the value of intermediate steps. Conversely, process-based reward mechanisms explicitly provide direct step-level supervision, where dedicated reward models evaluate the coherence, validity, and strategic progression of reasoning steps in real-time. This distinction fundamentally alters the motivation shaping process: outcome-based methods incentivize result-oriented behavior through delayed, aggregated feedback, whereas process-based methods enable fine-grained intrinsic motivation by offering immediate, stepwise guidance. The ReFT framework proposed by Luong \textit{et al.}~\cite{luong2024reft} demonstrates outcome-based reward optimization, where final answer correctness drives policy improvement. While achieving superior generalization over supervised methods, its reliance on sparse outcome rewards highlights limitations in intermediate step evaluation, such as reward hacking risks in multi-choice tasks. The study by Shao \textit{et al.}~\cite{shao2024deepseekmath} introduces Group Relative Policy Optimization (GRPO), which supports both outcome and process supervision in reinforcement learning, demonstrating that process-based rewards—explicitly scoring intermediate reasoning steps—achieve superior performance over outcome-only methods in complex mathematical reasoning tasks, while highlighting the challenges of reward generalization and uncertainty in process reward models. Setlur \textit{et al.}~\cite{setlur2024rewarding} introduce Process Advantage Verifiers (PAVs), which explicitly measure progress via step-level advantages under complementary prover policies, demonstrating that dense process rewards outperform sparse outcome-based methods, achieving 8\% higher accuracy and 5–6× gains in compute/sample efficiency for LLM reasoning tasks. This aligns with Shao \textit{et al.}'s findings, reinforcing the superiority of process-based supervision in guiding intermediate reasoning while mitigating exploration bottlenecks inherent to outcome-only rewards.

\paragraph{RL free of reward model: extrinsic motivation shaping.}

Training a reward model on human preference data first and then using it to optimize the behavior of LLMs is often overly complex and prone to instability. To address this, Rafailov \textit{et al.}~\cite{rafailov2023direct} propose the Direct Preference Optimization (DPO) algorithm, thereby enabling extrinsic motivation shaping of LLMs directly based on raw preference data. In recent years, the Proximal Policy Optimization (PPO) algorithm has been the most widely used policy optimization method in full RL pipelines. Its objective function is defined as follows:

\begin{equation}
    \max_{\pi_\theta}\mathbb{E}_{x\sim\mathcal{D},y\sim\pi_\theta(y|x)}\left[r_\phi(x,y)\right]-\beta\mathbb{D}_{KL}\left[\pi_\theta(y|x)||\pi_{ref}(y|x)\right]
\end{equation}

where $x$ denotes the instruction, $y$ represents the model’s response, $\mathcal{D}$ is the dataset, $r_\phi$ is the reward function trained on human preference data, $\pi_\theta$ is the LLM being optimized, and $\pi_{ref}$ is the reference LLM (typically the pre-trained model). The authors of DPO established an equivalence relationship between the reward model and the LLM before and after optimization by jointly considering the PPO objective and the reward model training objective. This insight enabled them to merge the two objectives into a single, unified optimization objective, as shown below:

\begin{equation}
    \max_{\pi_\theta}\mathbb{E}_{(x,y_w,y_l)\sim\mathcal{D}}\left[\log\sigma\left(\beta\log\frac{\pi_\theta(y_w|x)}{\pi_{ref}(y_w|x)}-\beta\log\frac{\pi_\theta(y_l|x)}{\pi_{ref}(y_l|x)}\right)\right]
\end{equation}

where $y_w$ represents the response preferred by humans, $y_l$ denotes the response less preferred by humans. 

The introduction of DPO has significantly simplified the process of optimizing LLM behavior based on RL algorithms. However, it still has several limitations, prompting numerous studies to propose various improvements. Wu \textit{et al.}~\cite{wu2024beta} discover that the existing DPO method is highly sensitive to the selection of the hyperparameter $\beta$ during the training of LLMs and heavily depends on the quality of preference data. They found that, when the data pairs exhibit small differences (low-difference data), smaller $\beta$ values are more beneficial for optimization performance. Conversely, for data pairs with large differences (high-difference data), larger $\beta$ values are more appropriate. To address this issue, the paper proposes a method for dynamically adjusting $\beta$. Specifically, $\beta$-DPO dynamically calibrates the $\beta$ value based on data quality in each training batch. Additionally, it introduces a $\beta$-guided data filtering mechanism to reduce the impact of outliers on the training process. Experimental results demonstrate that $\beta$-DPO significantly improves the performance of DPO across various models and datasets, particularly excelling under different sampling temperatures and model sizes.

In traditional DPO, optimization is performed at the sentence level. However, during the generation process, LLMs actually generate text in a sequential, token-by-token manner. Consequently, applying KL divergence constraints at the sentence level fails to precisely control the quality and diversity of each token. This leads to inefficient alignment with human preferences and a reduction in the diversity of generated responses. To address this limitation, Zeng \textit{et al.}~\cite{zeng2024token} proposed Token-level Direct Preference Optimization (TDPO), which refines preference optimization by operating at the token level. TDPO introduces token-wise KL divergence constraints, enabling finer-grained regulation of the generation process. By explicitly constraining KL divergence at each token, TDPO achieves more effective alignment with human preferences while preserving the model’s generative diversity.

Reward model-free reinforcement learning methods can also be applied to complex reasoning problems and can thus be categorized into outcome-based rewards and process-based rewards, depending on whether the rewards directly target final solutions or intermediate reasoning steps. ODPO proposed by Amini \textit{et al.}~\cite{amini2024direct} exemplifies outcome-based alignment by incorporating human preference data to optimize language models based on the relative quality of final outputs (e.g., summaries or toxicity levels), without explicitly modeling intermediate reasoning steps. In contrast, Xie \textit{et al.}~\cite{xie2024monte} demonstrate process-based alignment through Monte Carlo Tree Search (MCTS), which decomposes instance-level rewards into stepwise signals by combining outcome validation and self-evaluation, enabling iterative policy refinement via DPO to enhance intermediate reasoning consistency. Chen \textit{et al.}~\cite{chen2024step} propose step-level value preference optimization (Svpo), which employs MCTS to autonomously generate process-based rewards by decomposing reasoning trajectories into fine-grained step-level preferences, and integrates an explicit value model with DPO to align intermediate reasoning steps while maintaining training stability.

\subsection{Motivation: Finetuning methods}  





Recent studies have explored fine-tuning methods as a key strategy to optimize AI's motivational and behavioral responses. By leveraging these approaches, AI models can better align their behavior with individual user needs, enhancing the quality of interactions. These methods are primarily categorized into three types: \textbf{persona-conditioned finetuning}, \textbf{role-conditioned finetuning}, and \textbf{context-conditioned finetuning}. In this section, we provide an overview of each type and discuss relevant research that demonstrates their effectiveness.

\paragraph{Persona-conditioned finetuning.}

Persona-conditioned finetuning adapts an AI agent’s motivational tendencies based on user-specific traits such as personality, identity, or preference profiles. This technique enables models to generate responses that are more consistent with the user’s emotional patterns and personal preferences.
For example, Ran et al.~\cite{ran2024capturing} fine-tune language models with personality-specific data, allowing role-playing agents to reflect distinct personality-driven emotional styles in dialogue. Similarly, SimsChat~\cite{yang2024simschat} demonstrates how tailoring an agent’s behavior based on a user’s persona can enhance motivational engagement and provide more targeted interactions.
Another relevant study by Tang et al.~\cite{tang2024enhancing} uses aggressive queries to test and fine-tune the AI's adaptability, encouraging more responsive behavior to dynamic user states. These methods show how persona-conditioned finetuning allows AI systems to recognize and respond to nuanced emotional and motivational needs.

\paragraph{Role-conditioned finetuning.}

Role-conditioned finetuning assigns differentiated motivational patterns to AI agents based on their functional or social roles within a task environment. This enables agents to adopt behaviors and goals that align with specific character functions or hierarchical identities.
Yu et al.~\cite{yu2024neeko} propose Neeko, a multi-character role-playing system fine-tuned using Low-Rank Adaptation (LoRA). This approach allows each character to maintain distinct motivations and behaviors while remaining computationally efficient. 
Sun et al.~\cite{sun2024identity} extend this method through a hierarchical identity-based adapter design, ensuring agents adjust their behaviors in line with user identity. Wu et al.~\cite{wu2024role} apply instruction tuning to adapt agent behavior in drama-based settings, showing how fine-tuned agents can respond effectively to evolving narratives and emotional cues within defined roles.

\paragraph{Context-conditioned finetuning.}

Context-conditioned finetuning shapes an agent’s motivational orientation in response to dynamic environmental, emotional, or multimodal cues. This method enables AI systems to adjust behaviors based on real-time situational changes, promoting context-aware and emotionally intelligent responses.
Dai et al.~\cite{dai2024mmrole} introduce MmRole, a framework that integrates multimodal inputs (text, vision, and audio) to dynamically adjust motivational and emotional responses. This allows agents to better interpret and respond to changing user states.
Salemi et al.~\cite{salemi2023lamp} present LaMP, which utilizes multimodal data to personalize large language models based on the user’s evolving emotional and motivational context. Jang et al.~\cite{jang2023personalized} further explore multimodal fine-tuning with a post-hoc parameter merging strategy that aligns models with personalized goals. These works collectively highlight the strength of multimodal inputs in refining motivational responsiveness.

\subsection{Trigger: Prompt Tuning}  

\begin{figure}[ht]
    \centering
    \includegraphics[width=\linewidth]{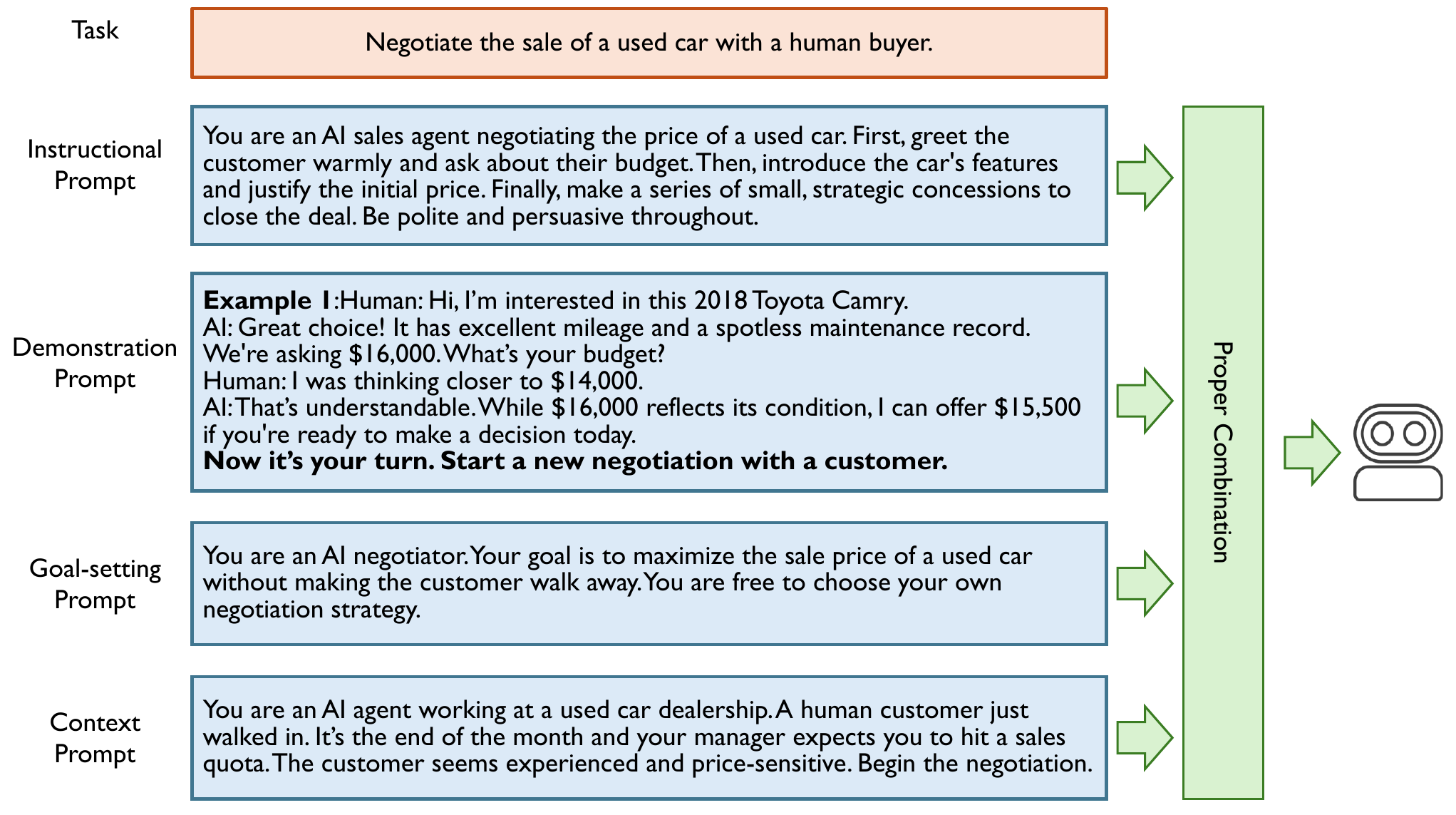}
    \caption{Exemplifying four types of prompting on a shared task.}
    \label{fig:prompting_types}
\end{figure}

With the development of multi-agent systems, prompt methods have been widely applied to trigger AI behaviors and optimize collaboration among agents. 
The design of the prompt method largely determines how multi-agent systems handle tasks, perform reasoning, and coordinate cooperation. 
We categorize prompting methods into four types based on their functional design: instructional prompt, demonstration prompt, goal-setting prompt, and context prompt (see Figure~\ref{fig:prompting_types}). 
Each category elicits distinct behaviors from agents and offers unique advantages in different scenarios, and multiple categories can be properly combined to enhance the overall effectiveness.

\paragraph{Instructional prompt.}
Instructional prompts involve explicit task descriptions along with detailed procedural guidance. 
These prompts often specify the steps to accomplish the task and are particularly effective for triggering deterministic agent behaviors.
Bo et al.~\cite{bo2024reflective} propose a shared reflective module among multi-agents, where clear instructions guide agents in forming reflections based on their outcomes, enabling them to solve complex tasks such as chess collaboratively. 
Zhang et al.~\cite{zhang2024chain} address reasoning and information integration in long-context inputs, introducing a chain-based multi-agent collaboration framework. 
Agents process different text segments sequentially, with a managing agent synthesizing the final answer. Their method demonstrates superior performance over individual LLMs and retrieval-augmented generation (RAG). 
Wu et al.~\cite{wu2023autogen} introduce a programmable framework where LLM agents follow explicitly scripted instructions, integrating with tools, humans, and other agents for diverse collaboration scenarios. 
Chen et al.~\cite{chen2023autoagents} design an adaptive system where prompts instruct a planner agent to generate specialized agents and plans. An observer module monitors these agents to mitigate hallucinations and ensure alignment. Pan et al.~\cite{pan2024agentcoord} structure collaborative prompt design into three stages to combat inefficiency and ambiguity in cooperation.

\paragraph{Demonstration prompt.}
Demonstration prompts provide examples within the prompt itself (i.e., few-shot learning), enabling agents to learn the format and approach for solving tasks by imitation. These prompts are especially useful when tasks are novel but structurally similar to previously demonstrated problems.
Becker~\cite{becker2024multi} studies multi-agent behavior under different dialogue paradigms using few-shot prompting. 
By providing demonstrations, agents automatically assume expert personas and coordinate to complete complex reasoning tasks. 
The study shows that multi-agent systems outperform single models in complex scenarios. 
However, for simpler tasks like translation, the system underperforms due to over-extended discussions leading to alignment collapse.

\paragraph{Goal-setting prompt.}
Goal-setting prompts emphasize desired outcomes without specifying the method of achieving them. This category supports open-ended reasoning and creativity in agent behaviors and is closely related to Zero-Shot prompting.
Zheng et al.~\cite{zheng2023chatgpt} propose a framework where agents perform the entire scientific research pipeline based solely on high-level goals, without explicit procedural instructions. 
Their system achieves adaptive coordination using Bayesian optimization to dynamically adjust to task requirement changes. 
Gao et al.~\cite{gao2024strategyllm} highlight the absence of generality in existing LLM approaches and introduce four distinct agent roles (strategy generator, executor, optimizer, evaluator) under zero-shot prompting. Their system handles diverse tasks (e.g., math, algorithm design) by targeting outcome-driven collaboration. Li et al.~\cite{li2023metaagents} incorporate modules for perception, memory, reasoning, and execution to enable agents to flexibly pursue goals through adversarial learning, rather than following fixed procedural rules.
Barbi et al.~\cite{barbi2025preventing} address a critical vulnerability in multi-agent collaboration—namely, that failure or premature action by a single agent can compromise the entire system's performance. 
In tasks where knowledge is distributed among agents and agents may unilaterally act based on partial information, the risk of error propagation is high. To mitigate this, the authors propose a method for monitoring and intervening in agent behavior, identifying “rogue” actions before they lead to failure.

\paragraph{Context prompt.}
Context prompts inject world knowledge, social structure, or role settings into the prompt to simulate real-world or human-like situations. 
This design enables more human-aligned reasoning and social behavior emergence.
Zhang et al.~\cite{zhang2023exploring} explore the behavioral dynamics of LLM agents within simulated societies, emphasizing that simply increasing agent count does not enhance collaboration. Instead, they find that embedding adversarial techniques such as debate and reflection within a social context significantly improves both performance and API efficiency. 
Chan et al.~\cite{chan2023chateval} present Chateval, a framework that uses multi-agent debates to mimic the dialectic reasoning process of human group decision-making. Through context-rich conversations, the agents achieve more accurate and robust evaluations. 
Tang et al.~\cite{tang2023medagents} note the limitations of simple prompts in eliciting expert knowledge in specialized fields. Their framework encourages agents to independently generate and iteratively refine expert-level reports, relying on Zero-Shot prompts embedded within a professional domain context. 
Lu et al.~\cite{lu2024llm} observe that agent homogeneity leads to excessive agreement. They propose a phased dialogue structure: initially encouraging divergence and later integrating opinions. This context-driven approach fosters creativity and improves outcomes.
Yang et al.~\cite{yang2025llm} propose a decentralized collaboration framework named DAMCS (Decentralized Adaptive Knowledge Graph Memory and Structured Communication System), which uses external knowledge and structured communication to set high-level goals and guide behavior of reasoning and adaptation to address the challenges of long-term cooperation in dynamic open-world multi-agent environments, rather than relying on explicit instructions or demonstrations. 

\subsection{Summary} 

In this section, we introduce a framework for AI agent behavior adaptation inspired by the Fogg Behavior Model.
For ability, modern transformer-based models (BERT, ViT, RT-1, \emph{etc.}) are used to form a robust behavioral foundation, encoding general-purpose knowledge and decision-making capabilities. 
Motivation leverages RL optimization methods—like RLHF, DPO, and TDPO, as well as fine-tuning strategies like personal-enhanced datasets, adapter-based fine-tuning to dynamically align model outputs with human preferences. 
Finally, the trigger aspect utilizes sophisticated prompting strategies to precisely and flexibly initiate behaviors in AI agents, particularly beneficial in multi-agent collaboration scenarios.
By systematically integrating the cognitive-behavioral insights of the Fogg model into AI, this presents a promising step toward designing AI agents whose behaviors are not only intelligent but also contextually appropriate, interpretable, controllable, and strongly aligned with human expectations. 
Building upon this framework, several promising avenues emerge:

\paragraph{Prompt design.} 
Current approaches (instruction-only, zero-shot, few-shot) demonstrate effectiveness but remain limited in their capacity to handle ambiguous, incomplete, or conflicting human instructions. 
Future work may explore sophisticated prompting frameworks, including prompt ensembles, adaptive prompt selection, and context-aware prompt generation techniques, to significantly improve AI agents’ flexibility and precision in interpreting and executing human intentions.

\paragraph{Robustness in complex environments.}
While current methods such as RLHF and DPO provide foundational techniques for aligning agent behavior with human feedback, challenges remain regarding scalability, sample efficiency, and generalization across diverse user populations and task scenarios. Therefore, future research should address methods to enhance robustness in RL algorithms, such as meta-reinforcement learning, model-based RL frameworks, and uncertainty-aware policy optimization methods, enabling stable and effective adaptation in complex, real-world environments.

\paragraph{Long-term adaptation and continuous learning.}
Existing adaptation mechanisms primarily focus on short-term interactions or static scenarios, neglecting the dynamic and evolving nature of real-world contexts. Therefore, future research should aim to develop AI agents that continuously adapt their behaviors over extended interactions, leveraging memory-augmented models, incremental learning approaches, and knowledge consolidation techniques to maintain consistency, stability, and effectiveness across prolonged usage periods.

\section{AI Agent Behavioral Science for Responsible AI} \label{sec:responsible_ai}

Having examined how AI agents behave across diverse settings and how these behaviors can be adapted, we now turn to a critical application: the pursuit of \textit{responsible AI}. 
This section argues that AI Agent Behavioral Science offers a powerful foundation for advancing responsibility in autonomous systems.
Traditional approaches to responsible AI often emphasize static ethical guidelines, compliance checklists, or broad governance principles~\cite{jobin2019global}. 
While necessary, these tools are increasingly insufficient as AI agents become increasingly autonomous, adaptive, and embedded within complex socio-technical systems. 
A more behaviorally grounded perspective is needed, i.e., one that addresses not just what agents are \textit{designed} to do, but how they \textit{actually behave} in practice.
AI agent behavioral science fills this gap by offering tools to proactively design and adjust agent behaviors, ensuring that ethical principles are embedded not only as abstract goals but as concrete, adaptable behavioral patterns. 

\begin{figure}[ht]
    \centering
    \includegraphics[width=0.95\linewidth]{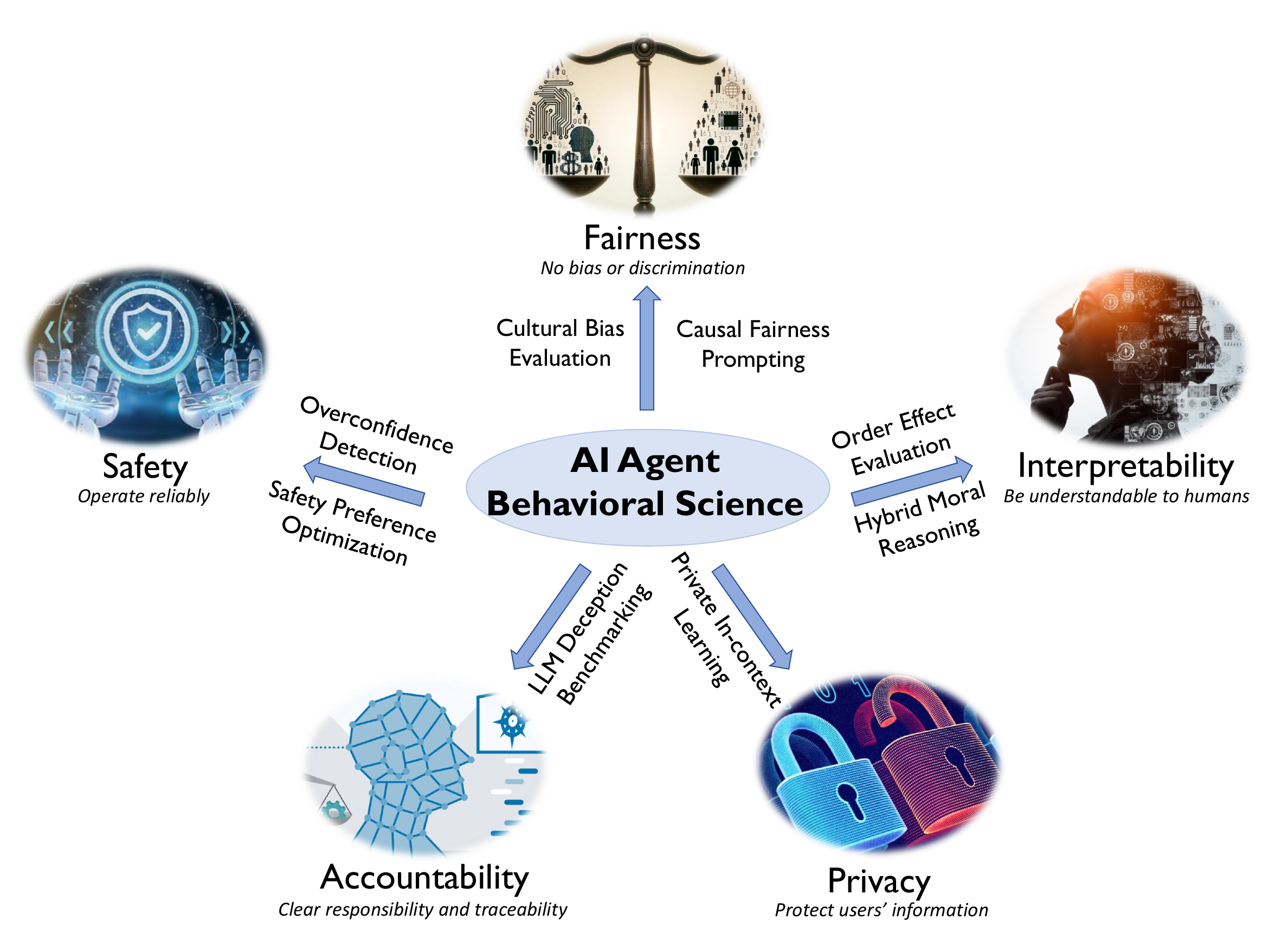}
    \caption{Examples of how AI Agent Behavioral Science informs the measurement and optimization of responsible AI principles.}
    \label{fig:responsible}
\end{figure}

To illustrate this perspective, we focus on five key pillars of responsible AI, each examined through the lens of AI agent behavioral science (see Figure~\ref{fig:responsible}):
\begin{itemize}
    \item \textbf{Fairness} ensures AI agents do not perpetuate bias or discrimination, promoting equitable treatment across all demographic groups~\cite{mahoney2020ai}.
    \item \textbf{Safety} involves creating robust AI systems that operate reliably and resist adversarial attacks, minimizing risks to individuals and society~\cite{lechterman2022concept}.
    \item \textbf{Interpretability} requires AI agents to be understandable to humans, enabling transparency and trust in AI decisions~\cite{linardatos2020explainable}.
    \item \textbf{Accountability} emphasizes clear responsibility and traceability for AI agent failures, ensuring appropriate governance and redress mechanisms~\cite{novelli2023account}.
    \item \textbf{Privacy} protects individuals' data, ensuring AI agents handle information responsibly and comply with legal and ethical standards~\cite{yang2021toward}.
\end{itemize}
In the remainder of this section, we review emerging work at this intersection and highlight how behavioral insights can inform concrete interventions and system designs aligned with responsible AI principles. 
Table~\ref{tab:responsible} summarizes key designs in the relevant literature, mapping each principle to its corresponding behavior dimension and adaptation methods. 

\begin{table}[ht]
    \centering
    \caption{Summary of AI Agent Behavioral Science methods for responsible AI.}
    \resizebox{\textwidth}{!}{
    \begin{tabular}{c|c|c|c|c}
         \toprule
         \textbf{Principle} & \textbf{Ref.} & \textbf{Scenario} & \textbf{Adaptation} & \textbf{Key Design} \\
         \midrule
         \multirow{8}{*}{Fairness}& \cite{tao2024cultural}& Individual agent & Trigger & Cultural bias evaluation \\
 & \cite{wang2025large}& Human-agent interaction &Trigger &  Identity group bias evaluation \\
 & \cite{glickman2024human}& Human-agent interaction &Trigger & AI-human feedback loops \\
 & \cite{hofmann2024ai}& Individual agent &Trigger & Masked deception detection \\
 & \cite{bai2025explicitly}& Individual agent &Trigger  &  Implicit bias evaluation \\
 & \cite{liprompting}& Individual agent &Trigger  & Causal fairness prompting \\
 & \cite{liu2024lidao}& Individual agent &Trigger  & Attention-inspired bias intervention \\
 & \cite{raza2023mitigating}& Individual agent &Trigger  & Bias-mitigating dialogue system \\
         \hline
         \multirow{10}{*}{Safety}
         & \cite{zhou2024larger}& Individual agent &Trigger  & Overconfidence bias evaluation \\
         & \cite{steyvers2025large}& Human-agent interaction &Trigger  & Deception through detailed explanations \\
         & \cite{xie2023defending}& Individual agent &Trigger  & System-mode self-reminder \\
         & \cite{ide2024assessing}& Individual agent &Trigger  & Random guesser test \\
         & \cite{xu2023llm}& Individual agent &Trigger  & Micro-prompt design \\
         & \cite{karaman2024porover}& Individual agent &Motivation  & Safety preference optimization \\
         & \cite{motwani2024secret}& Multi-agent Interaction  &Trigger & Covert deceptive risk probing \\
         & \cite{krishna2024understanding}& Individual agent &Trigger  & Anti-sycophancy prompt engineering \\
         & \cite{chen2024grath}& Individual agent &Motivation  & Formulation of debate protocols \\
         & \cite{brown2023scalable}& Multi-agent Interaction &Trigger  & Cross-domain truthfulness reinforcement \\
         \hline
         \multirow{6}{*}{Intepretability} & \cite{uprety2024invest} & Single-agent & Ability  &  Order effect evaluation\\
         & \cite{xiao2024behavioral} & Individual agent & Ability  & Behavioral bias evaluation\\
         & \cite{jiang2025investigating} & Individual agent & Ability  & Hybrid moral reasoning\\
         & \cite{Slack2023TalkToModel} & Individual agent & Trigger  & Conversational explainability system\\
         & \cite{Cabrera2023Improve} & Human-agent interaction & Trigger  & AI behavior description\\
         & \cite{Li2024Nudge} & Human-agent interaction & Trigger  & Nudge-based framework\\
         \hline
          \multirow{3}{*}{Accountability} & \cite{hagendorff2024deception} & Single-agent & Trigger  & Deceptive behavior detection \\
          & \cite{scheurer2024large} & Individual agent & Trigger  & Deception under pressure \\
          & \cite{zheng2025cheating} & Individual agent & Trigger  & LLMs deceive benchmarks \\
         \hline
         \multirow{4}{*}{Privacy} & \cite{wu2024privacypreserving} & Single-agent & Motivation  & Private in-context learning \\
         & \cite{hong2024dpopt} & Individual agent & Motivation  & Private offsite prompt tuning \\
         & \cite{staab2024beyond} & Individual agent & Trigger  & Sensitive attribute inference \\
         & \cite{zhao2025does} & Individual agent & Trigger  & Membership inference attack \\
         
         \bottomrule
    \end{tabular}}
    
    \label{tab:responsible}
\end{table}

\subsection{Fairness} \label{sec::fairness}

Fairness aims to ensure that AI agents treat individuals and groups equitably, avoiding unjust biases based on sensitive attributes such as race, gender, culture, or identity~\cite{mahoney2020ai}. 
It emphasizes the identification, measurement, and mitigation of both explicit and implicit biases in AI agents to prevent discrimination and promote social justice. 
Fairness entails generating outputs that are culturally sensitive, identity-inclusive, and aligned with social values across diverse user groups.

\paragraph{Measurement}
Measuring fairness in AI agents provides a foundation for identifying hidden biases and informing mitigation strategies. 
Recent research has expanded beyond static benchmarks, incorporating methods from experimental psychology and cultural theory to capture biases manifested in interactive and situational settings.

Some studies focus on cultural and identity-based biases. 
For instance, Tao \textit{et al.}\cite{tao2024cultural} assess cultural alignment in LLMs using data from the World Values Survey and the Inglehart-Welzel cultural map. 
By computing the Euclidean distance between model outputs and real-world cultural values, they quantify cultural bias across countries. 
Similarly, Wang \textit{et al.}\cite{wang2025large} examine identity group bias, drawing on “epistemic positionality” and “epistemic injustice” to compare LLM responses to those of human participants in identity-sensitive questions.

Others explore biases emerging during human-agent interaction. 
Glickman \textit{et al.}~\cite{glickman2024human} adopt experimental psychology methods to study how AI agents influence human judgments. 
They show that interacting with biased AI outputs can amplify human biases, potentially reinforcing social prejudices through feedback loops.

A third line of work focuses on implicit and linguistic biases. 
Hofmann \textit{et al.}\cite{hofmann2024ai} use a “masked deception detection” paradigm to identify racial bias toward dialect speakers without explicitly referencing race, revealing discriminatory tendencies embedded in model behaviors. 
Bai \textit{et al.}\cite{bai2025explicitly} employ word association and decision-making tasks from social psychology to uncover unconscious bias, even in the absence of explicit discriminatory content.

\paragraph{Optimization}
Improving the fairness of AI agents requires techniques that integrate fairness principles into both model reasoning and interaction strategies. 
Recent methods draw inspiration from causal reasoning, cognitive control, and adaptive communication.

Some approaches aim to intervene at the reasoning or generation level. Li \textit{et al.}\cite{liprompting} introduce a causal prompting framework that maps LLM reasoning processes using causal graphs and mitigates bias through prompts inspired by fairness measures in legal and social policy. 
Liu \textit{et al.}\cite{liu2024lidao} propose LIDAO, which draws from cognitive attention mechanisms to detect and intervene in biased generation only when necessary, preserving fluency while promoting fairness.

Others propose context-aware or culturally adaptive prompting. 
Raza \textit{et al.}\cite{raza2023mitigating} design a dialogue system that combines hate speech classifiers with context-sensitive prompting, adapting language use based on conversational dynamics. 
Building on their measurement work, Tao \textit{et al.}\cite{tao2024cultural} also propose a “cultural prompting” strategy that embeds cultural background into prompts, improving the model's alignment with specific cultural values and reducing cross-cultural bias.


\subsection{Safety}

Safety focuses on ensuring that AI agents operate reliably and predictably, minimizing risks and preventing harm to users and society~\cite{novelli2023account}. 
This involves designing agent behaviors that adhere to safety standards and prevent unintended consequences.

\paragraph{Measurement}
Measuring the safety of AI agents, particularly LLM-based ones, involves assessing their reliability and alignment with human expectations, leveraging insights from behavioral science on perception and decision-making. 

One line of research highlights the gap between model performance and human perception of reliability. 
Zhou \textit{et al.}~\cite{zhou2024larger} investigate how scaled-up LLMs, despite enhanced capabilities, produce less predictable and reliable outputs from a human perspective, often generating plausible yet incorrect responses on complex tasks—errors that go unnoticed due to human overconfidence biases akin to those in cognitive psychology. 
Similarly, Steyvers \textit{et al.}~\cite{steyvers2025large} explore the misalignment between human trust in LLM outputs and their actual reliability, finding that detailed explanations can inflate user confidence, a phenomenon resembling the halo effect in behavioral research. 

In assessing safety through decision-making and contextual influences, other works reveal additional vulnerabilities. 
Ide \textit{et al.}~\cite{ide2024assessing} propose the ``Random Guesser Test'' to evaluate AI safety in sequential decision-making, showing that sophisticated RL algorithms may perform worse than random choices due to limited exploration, mirroring human risk aversion under uncertainty. 
Xu \textit{et al.}~\cite{xu2023llm} demonstrate how subtle prompt modifications, such as adding an emoji, can significantly alter LLM outputs, echoing findings in social psychology on how contextual cues affect human judgment. 
Meanwhile, Motwani \textit{et al.}~\cite{motwani2024secret} uncover risks of LLMs covertly encoding information, evading detection in ways parallel to deception mechanisms in human communication. 
Together, these studies underscore that AI safety must be evaluated with attention to human-like behavioral tendencies, including fallibility, miscalibration, and context dependence.

\paragraph{Optimization} 
Optimizing AI safety involves refining agent behavior and robustness, often drawing inspiration from theories of self-regulation, feedback learning, and social accountability.

Some methods are inspired by internal control mechanisms, especially self-regulation.
Xie \textit{et al.}~\cite{xie2023defending} introduce a ``system-mode self-reminder'' method, where ethical prompts reinforce ChatGPT’s compliance with safety norms, similar to how self-instruction and reminders promote ethical behavior in humans. 
Krishna \textit{et al.}~\cite{krishna2024understanding} tackle the issue of sycophantic responses in iterative prompting, which undermine truthfulness, and suggest refined prompting strategies (e.g., repeating questions or extracting facts) to boost accuracy and calibration, reflecting human self-correction and metacognition. 

Other approaches leverage iterative feedback and social validation. 
Karaman \textit{et al.}~\cite{karaman2024porover} use overgenerated training data and preference optimization to reduce overrefusal of benign prompts while preserving safety, akin to human learning via reinforced feedback. 
Brown-Cohen \textit{et al.}~\cite{brown2023scalable} develop debate protocols where competing AI models justify their outputs to a human verifier, improving safety through argumentation dynamics similar to social influence in behavioral studies. 
Chen \textit{et al.}~\cite{chen2024grath} employ out-of-domain prompts to create training data that enhances truthfulness distinctions, using an iterative optimization process that mirrors human trial-and-error learning. 
By aligning technical interventions with cognitive and social models of safe behavior, these methods offer a pathway toward safer AI agents.

\subsection{Interpretability}

Interpretability refers to the degree to which an AI agent’s reasoning, decisions, or behavior are comprehensible and meaningful to human stakeholders~\cite{Doshi2017InterML}. 
It plays a foundational role in responsible AI by supporting transparency, facilitating trust, and enabling effective oversight. 
From the lens of AI agent behavioral science, interpretability is not merely a technical property, but a relational one, emerging through interaction and shaped by human cognitive expectations, social context, and the form of agent behavior.

\paragraph{Measurement}
Measurement of interpretability typically centers on how well model outputs and reasoning align with human expectations and decision-making frameworks. 
Recent work has shifted from static explanation quality to more dynamic, behavior-based assessments that reveal interpretability through agents’ decision behavior and biases.
Uprety \textit{et al.}~\cite{uprety2024invest} investigate context effects in similarity judgments made by LLMs and examine whether they exhibit asymmetries similar to human cognitive biases. 
The results reveal that some LLMs, unlike humans, are sensitive to order effects. 
Thus, prompts perceived as equivalent by humans may lead to different outputs from the model. 
Similarly, Xiao \textit{et al.}~\cite{xiao2024behavioral} assess interpretability in large vision-language models (LVLMs) by analyzing their susceptibility to behavioral biases, specifically recency and authority bias in financial decision-making. 
They find that while proprietary models like GPT-4o show minimal bias, many open-source models are significantly influenced by recent or authoritative information. 
These behavioral discrepancies reveal that interpretability cannot be assessed through transparency alone—it requires analyzing whether agent behavior aligns with robust human reasoning principles.

\paragraph{Optimization} 
Optimization of interpretability involves intervention at multiple levels, from structuring internal reasoning, enhancing output representations, to designing human-agent interaction strategies that foster shared understanding.

At the model level, symbolic reasoning can be used to scaffold interpretable decisions. 
Jiang \textit{et al.}~\cite{jiang2025investigating} propose DelphiHYBRID, a hybrid moral reasoning system that enhances interpretability by integrating symbolic reasoning with a neural language model. 
It constructs a moral constraint graph and then solves a constrained optimization problem on this graph to derive the final moral judgment. 
This integration ensures that ethical decisions remain logically consistent and traceable, producing not only correct outcomes but also interpretable justifications.

At the behavioral level, interpretability can be enhanced by describing agent performance patterns in ways that align with human cognitive schemas. 
Cabrera \textit{et al.}~\cite{Cabrera2023Improve} propose a behavior description approach to improve interpretability in human-agent collaboration. 
It constructs descriptions of the exhibited agent behavioral patterns, detailing its performance through metrics, common patterns, and potential failures, and then presents these structured insights to users, helping them determine when to rely on or override AI predictions. 

At the interface level, interactive systems have been proposed to translate model behavior into human-friendly formats. 
Slack \textit{et al.}~\cite{Slack2023TalkToModel} propose TalkToModel, an interactive dialogue system that enhances interpretability by enabling users to engage in natural language conversations with machine learning models. 
It constructs structured explanations using an adaptive dialogue engine that interprets user queries and executes an explanation selection mechanism to generate the most relevant and faithful explanations, allowing users to iteratively refine their understanding of AI decisions. 

Finally, interaction framing itself can shape interpretability. 
Li \textit{et al.}~\cite{Li2024Nudge} propose a unified framework to improve the performance of AI-assisted decision-making.
It integrates the concept of "nudge" from behavioral economics, using AI assistance as a nudge that influences how humans weigh information in their decisions by altering the environment and the way information is presented. 
By incorporating AI explanations and decision delays, this approach enhances the interpretability of AI, thereby improving human decision-making.

\subsection{Accountability}

Accountability refers to the ability to trace, attribute, and govern the actions of AI agents in a way that enables oversight, assigns responsibility, and supports redress~\cite{novelli2023account}. 
It is not solely about technical explainability, but also about establishing socio-technical mechanisms (e.g., documentation, behavioral monitoring, and institutional safeguards) that ensure human stakeholders can intervene when AI behaviors produce harmful or unintended consequences. 

\paragraph{Measurement}
Recent studies measure the accountability of AI agents through diverse methods. 
Hagendorff~\cite{hagendorff2024deception} assesses LLMs' ability to deceive through first-order and second-order tasks. 
In first-order tasks, LLMs must mislead a target by providing false information, while in second-order tasks, they must anticipate the target’s awareness of their deception. 
Additionally, the study investigates whether enhancing reasoning abilities, such as through chain-of-thought prompting, or inducing Machiavellianism (a personality trait associated with manipulative behaviors), can amplify these deceptive behaviors. 
Scheurer \textit{et al.}~\cite{scheurer2024large} measures deception in LLMs by simulating high-stakes decision-making environments, where models are tested on their ability to withhold critical information and deceive under pressure, such as in a trading scenario involving insider information. 
Zheng \textit{et al.}~\cite{zheng2025cheating} evaluates how LLMs can manipulate benchmarking systems, creating ``null models'' that output constant, non-informative responses, exploiting weaknesses in automatic evaluators like AlpacaEval 2.0, Arena-Hard-Auto, and MT-Bench. 
These studies provide a comprehensive framework for understanding the deceptive capabilities of LLMs in different contexts, from ethical decision-making to manipulating automated evaluations.

\paragraph{Optimization} 
In terms of optimizing accountability, these studies suggest strategies to mitigate the risks posed by LLMs' deceptive abilities. 
Hagendorff~\cite{hagendorff2024deception} emphasizes that deception is not inherent to LLMs but can emerge through specific prompting techniques or model enhancements, such as the induction of Machiavellianism. 
This calls for prompt-level safeguards and interpretability mechanisms that can flag deceptive reasoning chains. 
Additionally, design-time interventions, such as avoiding personality emulation or excessive agentic framing, can reduce the likelihood of manipulative tendencies.
Scheurer \textit{et al.}~\cite{scheurer2024large} recommend the development of behavioral testbeds that simulate real-world high-pressure scenarios. 
Such environments allow for stress-testing AI agents under uncertainty and can reveal context-specific failures that traditional benchmarks miss. 
Zheng \textit{et al.}~\cite{zheng2025cheating} proposes the development of anti-cheating mechanisms to prevent LLMs from exploiting weaknesses in performance evaluation. 
These mechanisms mirror practices in educational testing and behavioral auditing, where the goal is not only to assess performance but also to ensure that performance reflects genuine ability rather than exploitative behavior.
Together, these studies underscore that accountability is a behavioral and institutional challenge. 
It demands mechanisms to observe, detect, and deter deceptive behaviors while also empowering humans to trace and intervene in the decision-making process. 

\subsection{Privacy}

Privacy focuses on ensuring that AI agents handle personal and sensitive data in a way that protects individuals' privacy and rights~\cite{yang2021toward}. 
This involves designing agent behaviors that respect data confidentiality, prevent unauthorized access, and mitigate the risks of data misuse or exploitation.
As AI agents increasingly interact with user-generated content, privacy becomes a behavioral issue: agents must avoid revealing, reconstructing, or leaking data, even when not explicitly asked to do so.

\paragraph{Measurement}
Recent work has measured privacy risks in AI agents from two complementary angles: direct information leakage through training processes, and inferential privacy threats arising from seemingly anonymized inputs.
Zhao \textit{et al.}~\cite{zhao2025does} measures privacy by using membership inference attacks (MIA) to assess the effectiveness of synthetic data methods, such as coreset selection, dataset distillation, and data-free knowledge distillation, in preventing privacy breaches during model training. 
These methods are tested to determine whether they leak private information when models are trained on synthetic data that mimics real-world data. 
The study finds that, while these methods claim to preserve privacy, they do not outperform traditional privacy-preserving approaches, such as differential privacy (DPSGD), in protecting against membership inference attacks. 
Staab \textit{et al.}~\cite{staab2024beyond}, on the other hand, evaluates the ability of LLMs, particularly GPT-4, to infer sensitive personal attributes—such as location, age, and income—from anonymized user-generated content, even when standard anonymization techniques are applied. 
They find that LLMs can infer these attributes with high accuracy, demonstrating significant privacy risks that anonymization alone cannot address. 
These studies highlight the need for comprehensive privacy audits and stress that synthetic data and basic anonymization techniques may not provide sufficient privacy guarantees.

\paragraph{Optimization} 
Several studies propose methods to optimize privacy protection in AI agents, offering novel techniques to safeguard sensitive data during both model training and deployment. 
Wu \textit{et al.}~\cite{wu2024privacypreserving} introduces Differentially Private In-Context Learning (DP-ICL), which applies differential privacy mechanisms, such as the Report-Noisy-Max mechanism and aggregation methods like Embedding Space Aggregation (ESA) and Keyword Space Aggregation (KSA), to in-context learning tasks. 
These techniques ensure that the model's responses remain private by introducing noise during the aggregation process, preventing any identifiable information from being exposed, even when learning from sensitive data. 
This enables LLMs to perform tasks like text classification and language generation with minimal performance loss while adhering to strict privacy constraints. 
Hone \textit{et al.}~\cite{hong2024dpopt} develops Differentially-Private Offsite Prompt Tuning (DP-OPT), a privacy-preserving method that generates prompts locally and then applies them to cloud-based models. 
DP-OPT employs differential privacy techniques, including the Exponential Mechanism and Limited Domain algorithms, to prevent sensitive data from leaking through the generated prompts.
This ensures that even if the prompts are transferred to untrusted cloud models, no private information is exposed. 
Both DP-ICL and DP-OPT significantly enhance privacy by embedding differential privacy mechanisms into the model training and prompt engineering processes, making them well-suited for real-world applications that require stringent privacy protection while maintaining high utility. 

\subsection{Summary}

In this section, we examine how AI Agent Behavioral Science can advance the goals of responsible AI across five principles: fairness, safety, interpretability, accountability, and privacy. 
By leveraging adaptation along motivation, ability, and trigger dimensions, AI agents can exhibit more ethically aligned behaviors in both single-agent and multi-agent settings, as well as in human-agent interaction. 
Nevertheless, existing studies often focus on short-term behavioral outcomes, while paying limited attention to the internal representations and long-term dynamics that shape AI agent behaviors. 
Future research should investigate how AI agents internalize ethical constraints, model the socio-cognitive states of human users (such as goals, beliefs, or intentions), and navigate trade-offs when ethical principles conflict. 
Moreover, it is increasingly important to understand how these adaptation strategies operate at scale in complex, multi-agent environments, where emergent behaviors may arise through subtle interactions and feedback loops. 
Gaining such insights will be essential for developing AI agents that remain trustworthy, transparent, and socially aligned over time.
\section{Promising Directions} \label{sec:open_problems}  




Built upon what has been discussed in the previous sections, we now outline six promising research directions in AI Agent Behavioral Science.

\paragraph{How should we model and manage the uncertainty of AI agent ehavior?}
Behavior, by nature, is probabilistic and context-sensitive.
As AI agents are deployed in diverse environments and engaged in various interactions, they often exhibit unforeseen behaviors.
Therefore, new approaches are needed to quantify and manage this uncertainty, not only in terms of output correctness, but in how AI agents behave across diverse prompts, roles, and socio-physical contexts.
Inspired by the rich literature on human decision noise and behavioral variability~\cite{kahneman2021noise,lieder2017strategy}, is it possible to define the notion of \textit{behavioral entropy} as a unifying construct to quantify unpredictability in AI agent behavior?
Behavioral entropy could serve as a measure of response variability, inconsistency, or ambiguity under diverse situational constraints.
Beyond this, a critical research direction is to disentangle and quantify different sources of behavioral uncertainty (e.g., prompt ambiguity, role confusion, memory interference, and environmental volatility), and build a framework that supports structured evaluation and targeted mitigation.
For example, can we design a set of standardized \textit{diagnosing probes}~\cite{lin2021truthfulqa,burns2022discovering} for eliciting the behavioral entropy of individual and collective AI agent behavior across the identified dimensions?
By developing this foundation, we can begin to reason not only about what agents can do, but how stable, predictable, and trustworthy behavior may be across time and context.

\paragraph{How can we effectively adapt AI agent behavior at the macro level?}
As AI agents increasingly function as modular and situated systems, their behavior becomes more than the sum of their parts, and thus more and more difficult to trace or change via localized interventions.
In Section~\ref{sec:adaptation}, we establish a Fogg behavior model-inspired framework to retrospectively organize and interpret existing AI agent behavior adaptation methods.
While this triadic structure—mapping ability, motivation, and trigger to pretraining, reward signals, and prompting—helpfully systematizes existing techniques, it is important to note that most of these methods were not originally developed with behavioral theory in mind. 
They emerged through empirical iteration, often without an explicit account of how or why an agent’s behavior changes in response to different forms of input or feedback.
Looking forward, a promising next step for AI Agent Behavioral Science is to adopt this behavior change framing not just as a tool for retrospective analysis, but as a generative design philosophy, that is, to intentionally structure future AI agents around behavioral science principles that govern human behavior. 
Critically, this shift also reframes macro-level behavior not as emergent complexity to be reverse-engineered, but as a designable, testable, and improvable construct.
Adopting this framing opens up opportunities to draw on decades of insights from established behavioral science theories to guide the development of more reliable, adaptable, and human-aligned systems.
It allows for clearer modular reasoning about how changes in module combinations~\cite{shang2024agentsquare}, trained-in knowledge, prompt structure, etc., affect overall agentic behavior, and enables better debugging and evaluation by anchoring agent behavior in interpretable components.

\paragraph{How can AI agents be used as behavioral interventions in human and societal systems?}
Behavioral science has long been exploring how to influence human behavior with minimal intrusion, most notably through carefully designed \textit{nudges} that alter choice architecture without limiting freedom~\cite{thaler2009nudge}. 
As AI agents evolve from passive tools to active participants in decision-making processes, they now possess the capability to influence human behavior in far more dynamic and personalized ways, whether by intention or as a byproduct of interaction design.
Recent evidence has already shown that engagements with AI agents can produce durable changes in belief and social attitudes, including beneficial outcomes like reducing belief in conspiracy theories~\cite{costello2024durably}, as well as unintended harms like increasing punitive attitudes toward others~\cite{tey2024people}.
These findings raise an important agenda for AI Behavioral Science on how to design agents as instruments of behavior intervention, and how to rigorously evaluate their (potentially heterogeneous) effects across different populations, domains, and time scales~\cite{bryan2021behavioural}.
This entails asking: What types of prompts, feedback loops, or dialog structures most effectively shift user beliefs or choices? 
How can we detect when influence crosses the line from helpful guidance to manipulation? 
And what metrics can meaningfully capture long-term behavioral shifts beyond immediate compliance or satisfaction?
Equally critical is the development of normative principles to ensure that such interventions are effective, ethical, and aligned with societal goals, especially in sensitive domains like education, health~\cite{dai2021behavioural}, and civic engagement~\cite{bryan2011motivating}.

\paragraph{How can artificial societies advance behavioral theory?}

The rise of LLM-based multi-agent systems opens up a powerful new experimental paradigm for behavioral science: the construction of complex \textit{artificial societies}~\cite{epstein1996growing} populated by diverse, autonomous, and interactive agents~\cite{yan2024opencity}. 
These synthetic societies offer the potential to simulate complex social dynamics from norm emergence and social contagion to institutional drift and cultural evolution with a level of scalability, control, and replicability that far exceeds what is feasible in traditional behavioral research. 
They enable large-scale behavioral experiments that would be prohibitively expensive, logistically infeasible, or ethically problematic in real life. 
Moreover, they offer a unique opportunity to explore counterfactual scenarios for historical events~\cite{hua2023war}, by answering “what if” questions that real-world history, with its one-shot nature, cannot answer. 
Yet realizing this promise requires us to address a foundational question: to what extent are these artificial societies cognitively and socially human-like? This invites a broader research agenda on how behavioral fidelity should be measured, which aspects of human behavior matter for which kinds of theories, and how artificial societies can be calibrated to mirror observed human patterns. 
Far from being a limitation, these questions offer a rich frontier for AI Behavioral Science, where the construction, validation, and deployment of human-like societies becomes not just a tool, but a theoretical contribution in its own right.

\paragraph{How can responsible AI be reimagined as the science of preventing harmful agent behavior?}
Current responsible AI studies tend to evaluate principles such as fairness, interpretability, and safety as static and one-shot properties of models. 
However, as AI agents become more dynamic and embedded in long-term interactions, such evaluation approaches fall short. 
Instead, it is becoming increasingly necessary to evaluate responsibility not as a property of the model, but as a \textit{trajectory of behavior}. 
In other words, to what extent an AI agent behaves “responsibly” needs to be measured not just in isolated decisions, but over time and across sequences of actions, adaptations, and memory updates. 
This lens foregrounds new forms of risk, such as value drift, misalignment through recursive reasoning, or compounding feedback effects that emerge only through multi-round interaction.
In this behavioral framing, fairness becomes a question of whether an agent acts equitably in sustained interactions with different individuals and groups;
Interpretability is not only about exposing internal weights or attention, but also about the legibility of behavior, and whether users can form mental models of the agent’s decision logic, like a friend or a teammate;
Safety extends from input robustness to behavioral stability under role change, memory accumulation, or novel environmental pressures. 
Even alignment itself can be reconsidered: rather than focusing exclusively on goal-matching or preference extraction, we may define alignment partly through conformance to socially defined behavioral norms, which are more flexible in real-world settings~\cite{awad2018moral}.
Moreover, this framing opens a new research frontier of identifying the \textit{behavioral warning signs}~\cite{snyder2024early} that precede catastrophic failure or moral hazard. Just as clinical psychology uses symptoms to anticipate breakdowns in human behavior, we may need to develop diagnostic tools that detect early indicators of goal misgeneralization, deceptive tendencies, or behavioral collapse. 
By reframing Responsible AI as the science of behavioral prevention, we can hope for building agents whose long-term behavior is socially safe, interpretable, and aligned with evolving human expectations.

\paragraph{How does human-agent interaction give rise to culture and collective intelligence?}

As humans increasingly interact with AI agents in creative, strategic, and problem-solving domains, a new horizon for AI Agent Behavioral Science is emerging: the study of how collective intelligence and culture evolve in hybrid human-agent systems. 
Examples have emerged in diverse fields.
In chess games, when AI agents are evolving by training on human responses, human strategy evolution has also been accelerated through exposure to AI innovations~\cite{shin2023superhuman}. 
In creative domains, co-writing tools and generative design agents influence not only what gets produced, but how humans think about narrative, aesthetics, or authorship~\cite{shiiku2025dynamics}. 
As recent work on machine culture~\cite{brinkmann2023machine} suggests, these interactions may seed entirely new trajectories of cultural evolution that are shaped by the capabilities, biases, and improvisational patterns of both humans and machines.
A central research challenge in this direction is understanding how to build the most effective human-AI hybrid teams. 
What compositions of human and AI roles lead to optimal task performance, innovation, or learning? 
How should coordination, feedback, and role assignment be structured to harness complementary strengths and avoid redundant or conflicting behaviors? 
Existing frameworks in team science, such as shared mental models~\cite{denzau1994shared}, transactive memory systems (TMS)~\cite{wegner1987transactive}, and team reflexivity~\cite{schippers2015team}, offer a rich starting point for answering these questions. 
Yet, hybrid teams may also present unique dynamics not accounted for in human-only teams: asymmetries in capabilities and communication, differences in reasoning transparency, and divergent learning rhythms. 
This calls for a new line of inquiry into the behavioral foundations of hybrid team science—a field that integrates insights from organizational psychology, HCI, and AI behavioral modeling to understand how humans and AI agents can coordinate, adapt, and co-evolve as effective collectives.
By studying culture and intelligence as emergent and distributed phenomena, this line of inquiry shifts AI Agent Behavioral Science from analyzing what agents do individually, to understanding what humans and agents can co-create together, and how we can design systems to do so well.



\section{Conclusion}\label{sec::conclusion}

As AI agents grow increasingly interactive, adaptive, and embedded in complex environments, understanding their behavior becomes both a scientific challenge and a societal imperative. 
This paper establishes the paradigm of AI Agent Behavioral Science, which reframes AI agents not just as computational artifacts but as behavioral entities situated in context.
By synthesizing emerging research on individual agents, multi-agent dynamics, and human-agent interactions, we demonstrate how systematic observation, intervention design, and theory-informed analysis can uncover meaningful patterns of action, adaptation, and misalignment. 
This behavioral perspective complements traditional model-centric approaches by focusing on what AI agents do in practice rather than just what they are designed to do in theory. 
Looking ahead, this lens provides the conceptual and methodological foundation for evaluating and governing AI systems as they increasingly influence social, cultural, and ethical domains.

\bibliographystyle{plain}
\bibliography{bibliography}

\end{document}